\documentclass[10pt,letterpaper]{article}
\usepackage[top=0.85in,footskip=0.75in]{geometry}
\usepackage{amsmath,amssymb}
\usepackage{changepage}
\usepackage[utf8x]{inputenc}
\usepackage{textcomp,marvosym}
\usepackage{cite}
\usepackage{nameref,hyperref}
\usepackage{enumitem}
\usepackage[right]{lineno}

\usepackage{microtype}
\DisableLigatures[f]{encoding = *, family = * }
\usepackage[table]{xcolor}
\usepackage{array}
\newcolumntype{+}{!{\vrule width 2pt}}
\newlength\savedwidth

\newcommand\thickhline{\noalign{\global\savedwidth\arrayrulewidth\global\arrayrulewidth 2pt}%
\hline
\noalign{\global\arrayrulewidth\savedwidth}}

\usepackage[aboveskip=1pt,labelfont=bf,labelsep=period,justification=raggedright,singlelinecheck=off]{caption}

\makeatletter
\renewcommand{\@biblabel}[1]{\quad#1.}
\makeatother

\usepackage{lastpage,fancyhdr,graphicx}
\usepackage{epstopdf}
\usepackage{babel}
\usepackage{bbm}
\usepackage{bbold} 
\usepackage[textwidth=1.3in]{todonotes} 

\usepackage{color}

\newcommand{\tl}{\hat{t}}

\newcommand{\od}[2]{\frac{\mathrm{d}#1}{\mathrm{d}#2}}

\newcommand{\expect}[1]{\mathbb{E}\!\left[#1\right]}

\newcommand{\lrrund}[1]{\!\left( #1 \right)}
\newcommand{\lreckig}[1]{\!\left[ #1 \right]}

\usepackage{arydshln}

\begin{document}
\vspace*{0.2in}

\begin{flushleft}
{\Large
\textbf\newline{Mesoscopic description of hippocampal replay and metastability in spiking neural networks with short-term plasticity} 
}
\newline
\\
Bastian Pietras\textsuperscript{1,2,3},
Valentin Schmutz\textsuperscript{4},
Tilo Schwalger\textsuperscript{1,2*}
\\
\bigskip
\textbf{1} Institute for Mathematics, Technische Universit\"at Berlin, Berlin, Germany
\\
\textbf{2} Bernstein Center for Computational Neuroscience, Berlin, Germany
\\
\textbf{3} Department of Information and Communication Technologies, 
Universitat Pompeu Fabra, Barcelona, Spain
\\
\textbf{4} Brain Mind Institute, School of Computer and Communication Sciences and School of Life Sciences, \'Ecole Polytechnique F\'ed\'erale de Lausanne (EPFL), Lausanne, Switzerland
\\
\bigskip


* schwalger@math.tu-berlin.de

\end{flushleft}

\section*{Abstract}

Bottom-up models of functionally relevant patterns of neural activity provide an explicit link between neuronal dynamics and computation. A prime example of functional activity patterns are propagating bursts of place-cell activities called hippocampal replay, which is critical for memory consolidation.
The sudden and repeated occurrences of these burst states during ongoing neural activity suggest metastable neural circuit dynamics. As metastability has been attributed to noise and/or slow fatigue mechanisms, we propose a concise mesoscopic model which accounts for both. Crucially, our model is bottom-up: it is analytically derived from the dynamics of finite-size networks of Linear-Nonlinear Poisson neurons with short-term synaptic depression. As such, noise is explicitly linked to spiking noise and network size, and fatigue is explicitly linked to synaptic dynamics.
To derive the mesoscopic model, we first consider a homogeneous spiking neural network and follow the temporal coarse-graining approach of Gillespie to obtain a ``chemical Langevin equation'', which can be naturally interpreted as a stochastic neural mass model. The Langevin equation is computationally inexpensive to simulate and enables a thorough study of metastable dynamics in classical setups (population spikes and Up-Down-states dynamics) by means of phase-plane analysis. An extension of the Langevin equation for small network sizes is also presented. The stochastic neural mass model constitutes the basic component of our mesoscopic model for replay. We show that the mesoscopic model faithfully captures the statistical structure of individual replayed trajectories in microscopic simulations and in previously reported experimental data. Moreover,
compared to the deterministic Romani-Tsodyks model of place-cell dynamics, it exhibits a higher level of variability regarding order, direction and timing of replayed trajectories, which seems biologically more plausible and could be functionally desirable. 
This variability is the product of a new dynamical regime where metastability emerges from a complex interplay between finite-size fluctuations and local fatigue.

\section*{Author summary}
Cortical and hippocampal areas of rodents and monkeys often exhibit neural activities that are best described by sequences of re-occurring firing-rate patterns, so-called metastable states. Metastable neural population dynamics has been implicated in important sensory and cognitive functions such as neural coding, attention, expectation and decision-making. An intriguing example is hippocampal replay, i.e. short activity waves across place cells during sleep or rest which represent previous animal trajectories and are thought to be critical for memory consolidation. However, a mechanistic understanding of metastable dynamics in terms of neural circuit parameters such as network size and synaptic properties is largely missing. We derive a simple stochastic population model at the mesoscopic scale from an underlying biological neural network with dynamic synapses at the microscopic scale. This "bottom-up" derivation provides a unique link between emergent population dynamics and neural circuit parameters, thus enabling a systematic analysis of how metastability depends on neuron numbers as well as neuronal and synaptic parameters. Using the mesoscopic model, we discover a novel dynamical regime, where replay events are triggered by fluctuations in finite-size neural networks. This fluctuation-driven regime predicts a high level of variability in the occurrence of replay events that could be tested experimentally.




\section*{Introduction}
Metastable dynamics of neural populations is an important concept in computational neuroscience with increasing experimental evidence \cite{LaCFon19,BriYan22}. It is loosely defined as a sequence of recurring, discrete ``states'' of population activity that last much longer than the rapid, jump-like transitions between states (typically hundreds of milliseconds to several seconds). Sequences of metastable states have been frequently observed in cortical and hippocampal areas during task engagement as well as during spontaneous, ongoing activity and have been linked to various sensory and cognitive functions \cite{RabHue08,DurDec08}. These functions include the encoding of sensory stimuli \cite{AbeBer95,MazFon15} and internal representations of expectation \cite{MazLaC19} and attention \cite{EngSte16}. In these studies, the statistical properties of metastable neural activity  can often be  explained by hidden Markov models with a few latent states \cite{EngSte16,MazFon15}. However, more complex spatio-temporal activity patterns such as sequences of burst activity across hippocampal place cells during periods of both exploration and immobility (``replay of trajectories'') of an animal can also be regarded as metastable activity.  
In in-silico studies, metastable dynamics also emerges in networks of excitatory and inhibitory spiking neurons. This is the case for finite-size networks with clustered connectivity \cite{LitDoi12,MazFon15}, spatially-structured networks with slow fatigue processes for hippocampal replay \cite{EckBag22} and even for unstructured random connectivity in the inhibition-dominated regime \cite{TarBru17,KulKno22}. Network models exhibiting metastable dynamics have also been used to explain the stimulus-dependence of cortical variability \cite{LitDoi12}.

The mechanisms of metastable dynamics are often explained using heuristic population, or firing-rate, models.
These mechanisms can be roughly divided into two types: one in which transitions between metastable states are induced by fluctuations and another one in which transitions are induced by the deterministic part of the dynamics. In the first case, noise is essential for metastability because the noiseless dynamics would not exhibit spontaneous transitions. In contrast, in the second case, transitions also occur in the noiseless dynamics, while noise can still be useful to model variability of state durations. An important instance for the first type are multi-attractor models in the presence of noise, such as  noisy bistable models for perceptual rivalry \cite{MorRin07,ShpMor09,CaoPas16} and alternating Up-Down states \cite{HolTso06,ErmTer10,JerRox17}. Transitions correspond to noise-induced escapes from the basins of attraction. A popular instance of the second type are transiently stable states governed by a slow fatigue variable such as adaptation or synaptic depression. In these fast-slow systems, rapid transitions occur when quasi-stationary states of the fast subsystem (e.g. population activity) destabilize or vanish as the slow subsystem (fatigue process) evolves on a longer time scale. A prototypical example are relaxation oscillations, i.e. a (noisy) limit-cycle with a strong time-scale separation, used e.g. to model regular alternations between Up and Down states in spontaneous cortical activity \cite{MatSan12,LevBuz19}. A complex example of fatigue-induced metastability is the Romani-Tsodyks ring model for nonlocal events in place cells resembling the hippocampal ``replay'' dynamics \cite{RomTso15}. This deterministic ring model resides in a traveling-wave state (``non-local events'') or in a quiescent state depending on the spatial profile of a slow synaptic depression variable, which leads to a complex spatio-temporal activity pattern. We mention that there are also other mechanisms of metastability including noisy excitable dynamics \cite{LevBuz19} and deterministic motions between saddle points \cite{RabHue08}, also referred to as heteroclinic cycles or winnerless competition \cite{SelTsi03}. Looking at empirical data, however, it can be hard to distinguish different mechanisms, especially at high noise levels.



There has been much effort to infer the mechanism underlying metastable dynamics by studying the consistency of experimental data with heuristic population models. For example, in the case of cortical and hippocampal Up and Down states \cite{MatSan12,JerRox17,LevBuz19} and for perceptual bistability \cite{MorRin07,ShpMor09,CaoPas16}, it has been suggested that populations models, where noise-induced transitions are modulated by a slow fatigue variable, are most consistent with the data. An important question that has received relatively little attention is whether such conclusions are also consistent with the underlying circuit properties at the microscopic scale, modeled as networks of spiking neurons with biologically realistic neuronal, synaptic and network properties. 
Unfortunately, a clear link between the employed population models and microscopic circuit models is largely missing, and it thus remains unclear how the mechanisms of metastability depend on physiological parameters.
While neuronal and synaptic properties can be accounted for by mean-field models of integrate-and-fire networks \cite{TsoPaw98, MazFon15}, the dependence of metastable dynamics on the number of neurons in the network is poorly understood. This latter aspect is particularly crucial in the context of metastability because fluctuations due to a finite number of neurons have been found to be essential for fluctuation-induced metastability by several detailed simulation studies \cite{LitDoi12,MazFon15} as well as theoretical considerations \cite{Bre10}. The description of these internally generated fluctuations requires population models at the \emph{mesoscopic} scale, where the finite network size is explicitly taken into account \cite{MatGiu02,SchDeg17,SchChi19}.
Previous models for determining the mechanisms of metastability cannot describe this dependence:
In heuristic population models, fluctuations were introduced ad hoc by adding a phenomenological noise term without a link to the network size. In the case of  mean-field models,  fluctuations are usually not described at all because they vanish in the mean-field limit of infinitely many neurons.

In this contribution, we develop a theoretical framework for mesoscopic population dynamics with slow fatigue that can describe metastable dynamics and links to an underlying microscopic description. To this end, we use a bottom-up approach starting from a finite-size network of linear-nonlinear Poisson (LNP) spiking neurons connected via dynamic synapses undergoing short-term plasticity (STP). From this microscopic model we derive stochastic differential equations for a few mesoscopic variables describing the coarse-grained population dynamics. We focus on STP in the form of short-term synaptic depression as a slow fatigue mechanism because it is a ubiquitous feature of neural networks in the brain \cite{AbbVar97,ZucReg02,CooSch03,HigCon06,OswUrb12} and has been implicated in important functions such as temporal filtering \cite{MerLin10,DroSch13}, multistability \cite{MonHan2012} and working memory \cite{MonBar08}.
Mean-field models of STP \cite{TsoPaw98} have recently gained renewed attention  \cite{TahTor20,GasKno21} in the context of the Montbri{\'o}-Paz{\'o}-Roxin theory for quadratic integrate-and-fire neurons \cite{MonPaz15,PieDev19}. 
However, in these models the mean-field description of STP is heuristic -- it is not derived from a microscopic model but introduced ad hoc at the population level. More importantly, the models are deterministic corresponding to the limit of infinitely many neurons, and thus cannot explain fluctuation-induced transitions among metastable states in finite-size networks. Recently, we have developed a mesoscopic bottom-up model for finite-size networks with STP, and have demonstrated that the mesoscopic model accurately reproduces the metastable Up-and-Down-states dynamics of the microscopic model \cite{SchGer20}. The mathematical structure of that model has the intricate form of a state-dependent doubly-stochastic point-process driving a system of stochastic differential equations. As such, it is difficult to analyze and it lacks a straightforward, efficient simulation algorithm. However, the mesoscopic theory of \cite{SchGer20} can be used as a starting point to derive a temporally coarse-grained stochastic dynamics in the form of a simple jump-diffusion process. For the case of synaptic depression and large network size, we also present a short direct derivation of the diffusion limit that yields a mesoscopic model in the form a simple diffusion process. 

As we shall show below, our bottom-up modeling framework for mesoscopic population dynamics permits a re-evaluation of existing heuristic models for metastability in terms of an underlying microscopic network model. As a first example, we consider a single population of excitatory neurons with synaptic depression that generates population spikes and can transition between Up and Down states. The corresponding mesoscopic population model is similar to the model by Holcman, Tsodyks and co-workers \cite{BarBao05,HolTso06,DaoDuc15}, which successfully reproduced experimental observations \cite{DaoDuc15}. The important difference is that in our mesoscopic model all parameters are fixed by the microscopic parameters. Thanks to the low-dimensional character of the reduced mesoscopic system, we can apply phase-plane analysis to study the emergence of multiple stable states that soon become metastable when decreasing the network size. 
As a second, more complex example, we revisit the Romani-Tsodyks (RT) model for hippocampal replay activity in circular networks of place cells with synaptic depression \cite{RomTso15}. We propose a spiking-neural-network implementation of the original firing-rate model. The corresponding mesoscopic population model with finite-size noise
enables us to shed new light on the mechanisms underlying hippocampal replay in place cells of area CA3 in the hippocampus. In the deterministic (and heuristic) RT model, irregular switching between metastable traveling waves of sequential neural activity and a quiescent state is solely controlled by local synaptic depression as a slow fatigue mechanism \cite{RomTso15}. Yet, it is unclear whether such metastable replay dynamics also occurs in finite-size networks of spiking neurons and whether, in this case, replay sequences are fatigue-induced (like in the RT model) or may also be driven by finite-size fluctuations. Our model being reduced from a microscopic network of a finite number of neurons, we can interpolate 
between a fatigue-induced regime and a novel regime of fluctuation-induced hippocampal replay. We show that these two regimes lead to very distinct statistical predictions, which can be tested experimentally.

The present paper is organized in three main parts. In the first part, we present the mesoscopic bottom-up model
in two variants, a diffusion and a jump-diffusion model. In the second part, we use a single population model to demonstrate the performance of the two variants with respect to network size. In the third part, we turn to the more complex scenario of metastable nonlocal replay events in hippocampal place cells. We compare a novel dynamical regime of fluctuation-induced replay with the deterministic replay dynamics of \cite{RomTso15}. In the Discussion, we explicate biological limitations of the model and possible extensions to address these limitations. We also discuss potential advantages of the novel fluctuation-induced replay dynamics. Finally, in the Methods section, we provide the derivation of the mesoscopic model as well as the details on numerical simulations and statistical procedures.

\section*{Results}

\subsection*{Mesoscopic description of microscopic network dynamics}
\label{sec:micro}

We study the dynamics of a network of $N$ spiking neurons that, on the microscopic level, are modeled as linear-nonlinear-Poisson (LNP) neurons \cite{Chi01,SimPan04,OstBru11,GerKis14} with dynamic synapses \cite{SchGer20,GalLoe20}.
 The synaptic dynamics is given by the Tsodyks-Markram model of short-term plasticity (STP) \cite{TsoPaw98}. For simplicity, we focus in the Results part on the special case where the synaptic dynamics corresponds to pure depression \cite{PfiDay09} and the linear filter of the LNP neurons has an exponential impulse response function \cite{OstBru11}. The general theory for the full Tsodyks-Markram model with depression and facilitation as well as the straightforward extension to general linear filters, enabling biologically more realistic neuronal dynamics \cite{OstBru11}, is provided  in the Methods part. 

For the case of synaptic depression and exponential impulse response function, the LNP-STP model is given by the stochastic differential equations
\begin{subequations}\label{eq:micro}
\begin{align}
  \od{h_i}{t}&=\frac{\mu(t)-h_i}{\tau}+\frac{JU_0}{N}\sum_{j=1}^Nx_j(t^-)s_j(t),\label{eq:micro_h}\\
  \od{x_i}{t}&=\frac{1-x_i}{\tau_D}-U_0x_i(t^-)s_i(t), \label{eq:micro_x}\\
  s_i(t)&=\frac{dn_i(t)}{dt}=\sum_{k}\delta(t-t_k^i),\qquad dn_i(t)\sim\text{Pois}\lreckig{f(h_i(t^-))dt},
\end{align}
\end{subequations}
for $i=1, \dots, N$. Here, $s_i(t)$ is the spike train of neuron $i$ with conditional intensities $\{f(h_i(t^-))\}_{i=1}^N$, $\mu(t)$ represents a common external current mimicking, e.g., feedforward input from other areas, and $\tau$ can be interpreted as the membrane time constant. The synaptic parameters are given by the overall synaptic weight factor $J$, the relative depletion of neurotransmitter by a single transmitted spike $U_0$ and the time scale of synaptic depression $\tau_D$. The variables $h_i$ and $x_j$ can be interpreted as the input potential of neuron $i$ and the availability of synaptic resources at the outgoing synapses of neuron $j$, respectively. The trajectories $h_i(t)$ are càdlàg and $h_i(t^-)$ denotes the left limit (the same holds for $x_i(t)$). It can also be noted that  $h_i(t) = h_j(t)$ for all $t\geq 0$ and for all $i$ and $j$ if at time $0$ all the $h_i$ share the same initial condition.



\begin{figure}[t]
\centering{
\includegraphics[width=0.9\columnwidth]{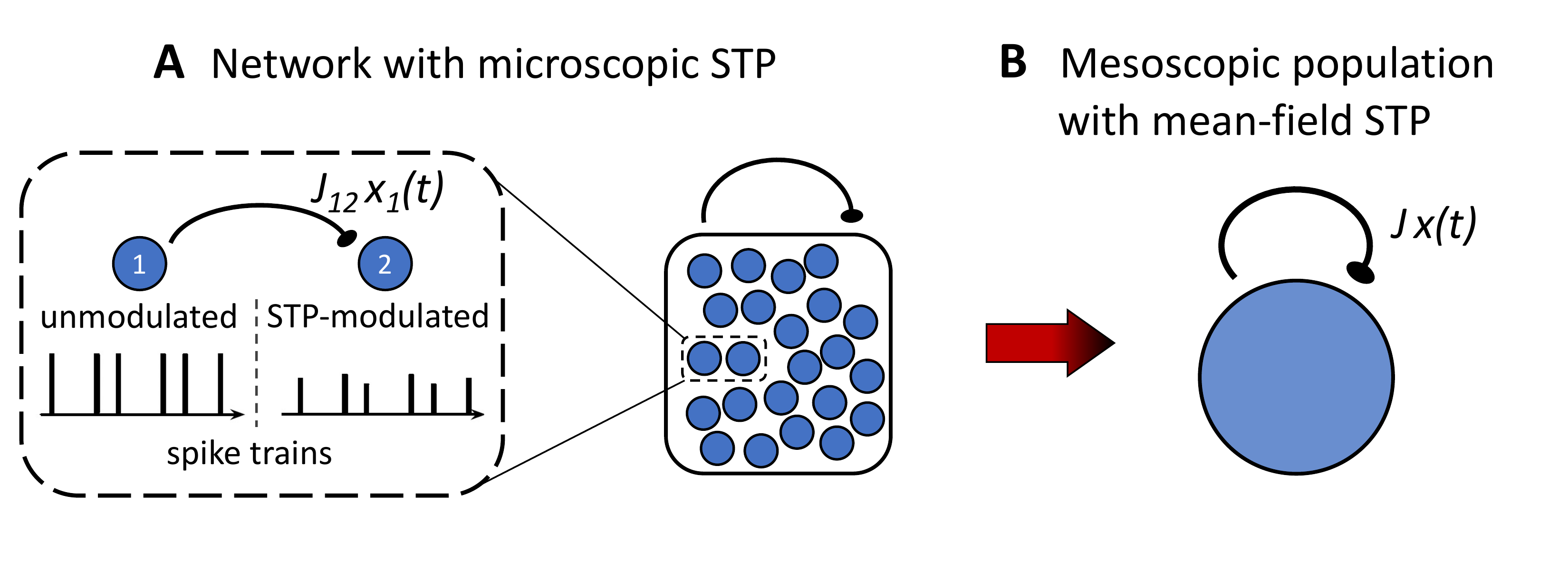}
\caption{{\bf From microscopic to mesoscopic population dynamics.} 
(A) Network with microscopic short-term plasticity. Dashed region shows a zoom into a pair of interconnected neurons: presynaptic neuron $1$ sends out an unmodulated spike train to postsynaptic neuron $2$ that receives the spike-train modulated by short-term depression.
(B) Mesoscopic mean-field model with one effective synapse undergoing short-term depression.} 
\label{fig0}}
\end{figure}

\subsubsection*{Diffusion model of the mesoscopic dynamics (Gaussian noise)}
Our goal is to derive a mean-field model for the microscopic dynamics Eq.~\eqref{eq:micro} that accounts both for the finite number of neurons as well as for the dynamic synapses undergoing Tsodyks-Markram STP, see Fig.~\ref{fig0}. The mean-field description will be based on the dynamics of the following mesoscopic variables defined as the empirical averages
\begin{equation} \label{eq:empirical}
    h(t):=\frac{1}{N}\sum_{i=1}^N h_i(t), \quad x(t):=\frac{1}{N}\sum_{i=1}^N x_i(t), \quad \text{and} \quad Q(t) := \frac{1}{N}\sum_{i=1}^N x_i^2(t).
\end{equation}
The desired dynamics of $h(t), x(t)$ and $Q(t)$ are supposed to no longer depend on (the index $i$ of) individual neurons, so we will approximate terms such as, e.g., the sum $\frac{1}{N}\sum_{i=1}^N x_i(t^-)s_i(t)$, by a diffusion term which only involves the mesoscopic variables.
To this end, we follow the temporal coarse-graining approach by Gillespie \cite{Gil00} for the derivation of a ``chemical Langevin equation'', see the Methods section for a detailed derivation.
In brief, we first use a \textit{macroscopically infinitesimal} time step $\Delta t$ \cite{Gil00} and approximate the coarse-grained sum $\int_t^{t+\Delta t}\frac{1}{N}\sum_{i=1}^N x_i(t^-)s_i(t)\,dt$ by a Gaussian random variable with variance proportional to $Q(t^-)$. In a second step, we derive the dynamics of $Q(t)$, discarding the fluctuations  whose effect on $h(t)$ and $x(t)$ is of order $N^{-3/2}$.
The resulting mesoscopic mean-field dynamics is given by the \textit{diffusion model}
\begin{subequations}\label{eq:meso}
\begin{align}
    \od{h}{t}&=\frac{\mu(t)-h}{\tau}+JU_0xf(h) + JU_0 \sqrt{\frac{Qf(h)}{N}}\xi(t), \label{eq:diff-model-h}\\
    \od{x}{t}&=\frac{1-x}{\tau_D}-U_0xf(h) - U_0\sqrt{\frac{Qf(h)}{N}}\xi(t),\\
    \od{Q}{t} &= 2\frac{x - Q}{\tau_D} - U_0(2-U_0)Qf(h),
\end{align}
\end{subequations}
where $\xi(t)$ is a Gaussian white noise with auto-correlation function $\langle\xi(t)\xi(s)\rangle=\delta(t-s)$. 
Although it is possible to deduce Eq.~\eqref{eq:meso} from the detailed doubly-stochastic mesoscopic dynamics derived in \cite{SchGer20} (see Methods ``Mesoscopic Tsodyks-Markram model [...]''), the derivation summarized above and presented in Methods ``Diffusion approximation ...'' is much simpler as it relies on a direct application of the diffusion approximation (avoiding the detour via the model presented in \cite{SchGer20}).

\subsubsection*{Jump-diffusion model of the mesoscopic dynamics (Hybrid noise)}
In large networks, it is plausible to assume that the spike input through a large number of recurrent connections can be approximated by a Gaussian process, and the diffusion model Eq.~\eqref{eq:meso} is valid for sufficiently large $N$.
In smaller networks, by contrast, we may no longer rely on the diffusion approximation since we need to take into account the shot noise character of the spike input. To this end, we start from the mesoscopic model of \cite{SchGer20} and derive a mesoscopic jump-diffusion model with facilitation and depression (see Methods). In this model, the noise takes on a hybrid form combining Poisson shot noise and Gaussian white noise. In the special case of short-term synaptic depression only, the resulting \textit{jump-diffusion model} of the mesoscopic dynamics reads
\begin{subequations}
\label{eq:model-depress-3}
\begin{align}
  \od{h}{t}&=\frac{\mu(t)-h}{\tau}+JU_0\Big[x(t^-)A(t)+\sqrt{\frac{\tilde Q f(h)}{N}}\xi_x(t)\Big],\label{eq:jump-diff-depress-h}\\
  \od{x}{t}&=\frac{1-x}{\tau_D}-U_0\Big[x(t^-)A(t)+\sqrt{\frac{\tilde Q f(h)}{N}}\xi_x(t)\Big],\label{eq:tilo_x}\\
  \od{\tilde Q}{t}&= - \Big[ \frac{2}{\tau_D} + U_0(2-U_0) f(h) \Big] \tilde Q + U_0^2 x^2 f(h).
\end{align}
Here, $\xi_x(t)$ is a Gaussian white noise with auto-correlation function $\langle\xi_x(t)\xi_x(s)\rangle=\delta(t-s)$ and
\begin{equation}
  \label{eq:popact}
  A(t)=\frac{1}{N}\frac{dn(t)}{dt}=\frac{1}{N}\sum_k\delta(t-t_k),\qquad dn(t)\sim\text{Pois}\left[Nf(h(t^-))dt\right],
\end{equation}
\end{subequations}
is a shot noise. The shot noise $A(t)$ is defined by the counting process $n(t)$ with jump times $t_k$ that occur with conditional intensity $Nf(h(t^-))$. The increment of the counting process $dn(t)$ represents the total number of spikes generated by all neurons in the small time interval $[t,t+dt)$, and $A(t)$ is therefore the empirical population activity. The presence of two different sources of noise in Eq.~\eqref{eq:model-depress-3} can be interpreted as the effect of two components that make up the synaptic input $N^{-1}\sum_ix_i(t^-)s_i(t)$ on the mesoscopic scale: First, a term $\lrrund{N^{-1}\sum_ix_i(t^-)}\cdot\lrrund{N^{-1}\sum_is_i(t)}=x(t^-)A(t)$ that arises if the variability of the weighting factors $x_i$ across synapses is neglected. This term represents the common spiking noise caused by shared spike inputs. Second, a correction term that accounts for the variability of $x_i$, approximated by a Gaussian distribution with variance $\tilde{Q}(t)$ as shown previously \cite{SchGer20}. 

Mathematically, the mesoscopic model, Eq.~\eqref{eq:model-depress-3}, is a jump-diffusion process because the shot noise leads to small jumps of order $1/N$ in addition to the diffusive dynamics caused by the Gaussian white noise. The jumps, however, occur at a high rate $Nf(h(t))$ so that in simulations with a coarse-grained time step $\Delta t$, unitary jumps will not be resolved. Instead, the increment of the spike count $\Delta n(t)=n(t+\Delta t)-n(t)$ can be drawn from a Poisson distribution with mean $Nf(h(t))\Delta t$ provided  a sufficiently small simulation time step $\Delta t\ll 1/f(h),\tau,\tau_D$.

We expect that the jump-diffusion model Eq.~\eqref{eq:model-depress-3} remains valid for small network sizes, for which the diffusion model Eq.~\eqref{eq:meso} ceases to provide an accurate description of the microscopic network dynamics Eq.~\eqref{eq:micro}.
In the large $N$-limit, the jump-diffusion model, Eq.~\eqref{eq:model-depress-3},  converges to the diffusion model, Eq.~\eqref{eq:meso}. In fact, by invoking the diffusion approximation, we can replace the shot noise by $A(t) \approx f(h) + \sqrt{f(h)/N}\xi_A(t)$, where $\xi_A(t)$ is an independent Gaussian white noise with auto-correlation function $\langle\xi_A(t)\xi_A(s)\rangle=\delta(t-s)$. As detailed in the Methods section, by combining the two independent noise terms $\xi_x(t)$ and $\xi_A(t)$ into a single noise process and identifying $\tilde Q$ with $Q - x^2$, we recover the diffusion model Eq.~\eqref{eq:meso} for large $N$.

\subsection*{Microscopic vs.\ mesoscopic dynamics of a single population exhibiting metastability}

The mesoscopic descriptions Eqs.~\eqref{eq:meso} and \eqref{eq:model-depress-3} of the full network of $N$ interacting spiking neurons with short-term depression (STD) effectively reduce the high-dimensional microscopic dynamics Eq.~\eqref{eq:micro} to a system of three stochastic differential equations in $h,x$, and $Q$.
In the limit $N\to \infty$, finite-size fluctuations in the mesoscopic dynamics vanish and the variable $Q$ becomes superfluous. The resulting two-dimensional macroscopic dynamics
readily allows for a comprehensive phase-plane analysis, see, e.g., \cite{HolTso06,DaoDuc15}, which reveals the deterministic backbone of, and can therefore yield important insights about, the full network dynamics.

To demonstrate the high accuracy of our mesoscopic description and also its usefulness for studying the effect of finite-size fluctuations on metastable dynamics, we will focus in this Section on two traditional examples of metastability in a single excitatory population ($J>0$) of LNP spiking neurons with STD: populations spikes and spontaneous transitions between Up and Down states.
For simplicity, we will assume in the following that the transfer function $f(h)$ has the form
\begin{equation}\label{eq:transfer}
    f(h) = r a \ln \big\{ 1 + \exp[ (h-h_0) / a ]\big\}
\end{equation}
with slope parameter $r$, smoothness $a$, and threshold $h_0$.
The transfer function $f(h)$ has an exponential sub-threshold tail and a linear supra-threshold part.
In the limit $a\to 0$, $f(h) = r[h-h_0]^+$ becomes a threshold linear function with slope $r$ and threshold $h_0$. The larger $a > 0$, the smoother the transition at the threshold.

\subsubsection*{Population spikes in an excitatory population}

\begin{figure}[!t]
    \includegraphics[width=.66\columnwidth]{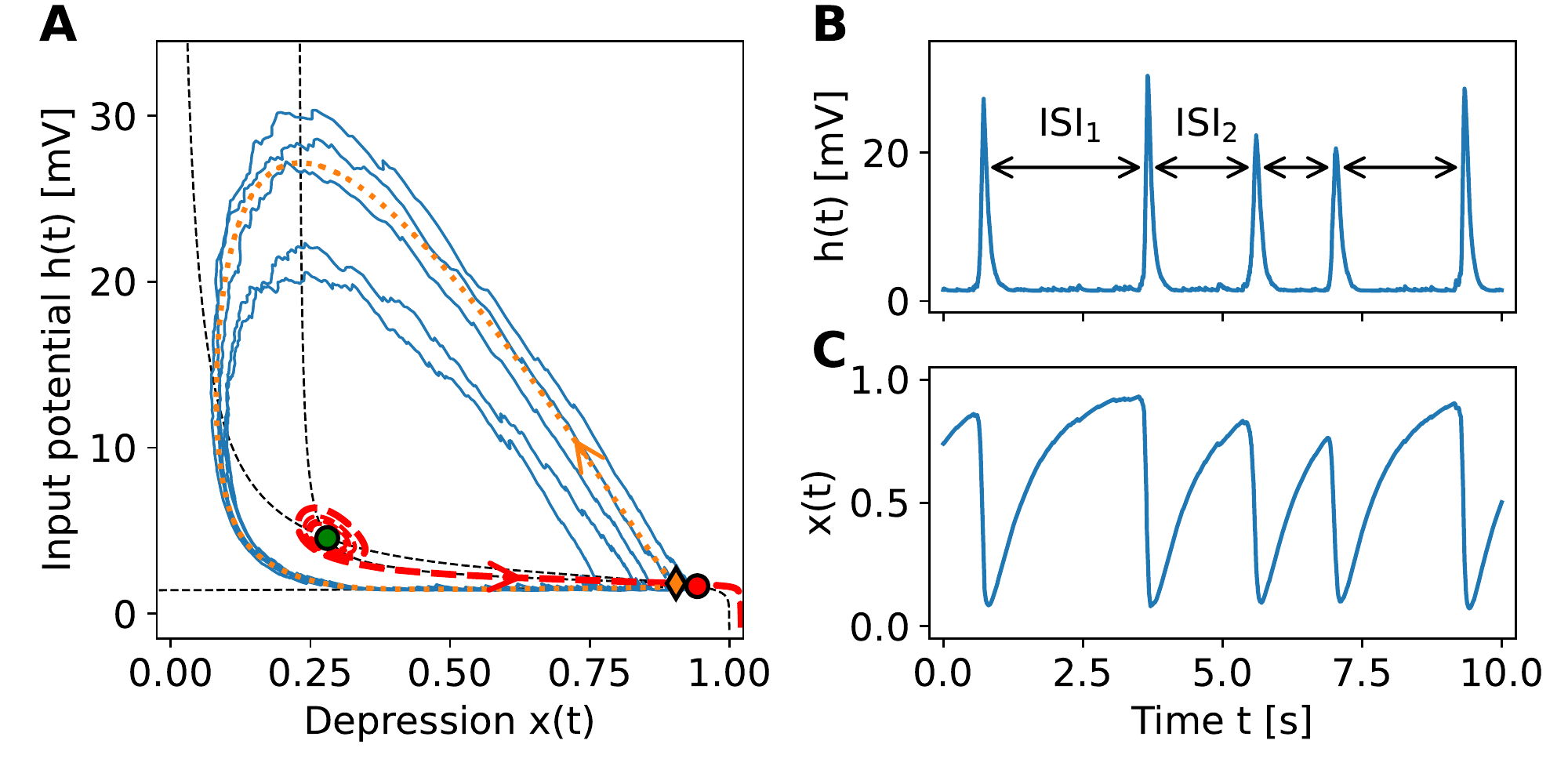}\\
    \includegraphics[width=\columnwidth]{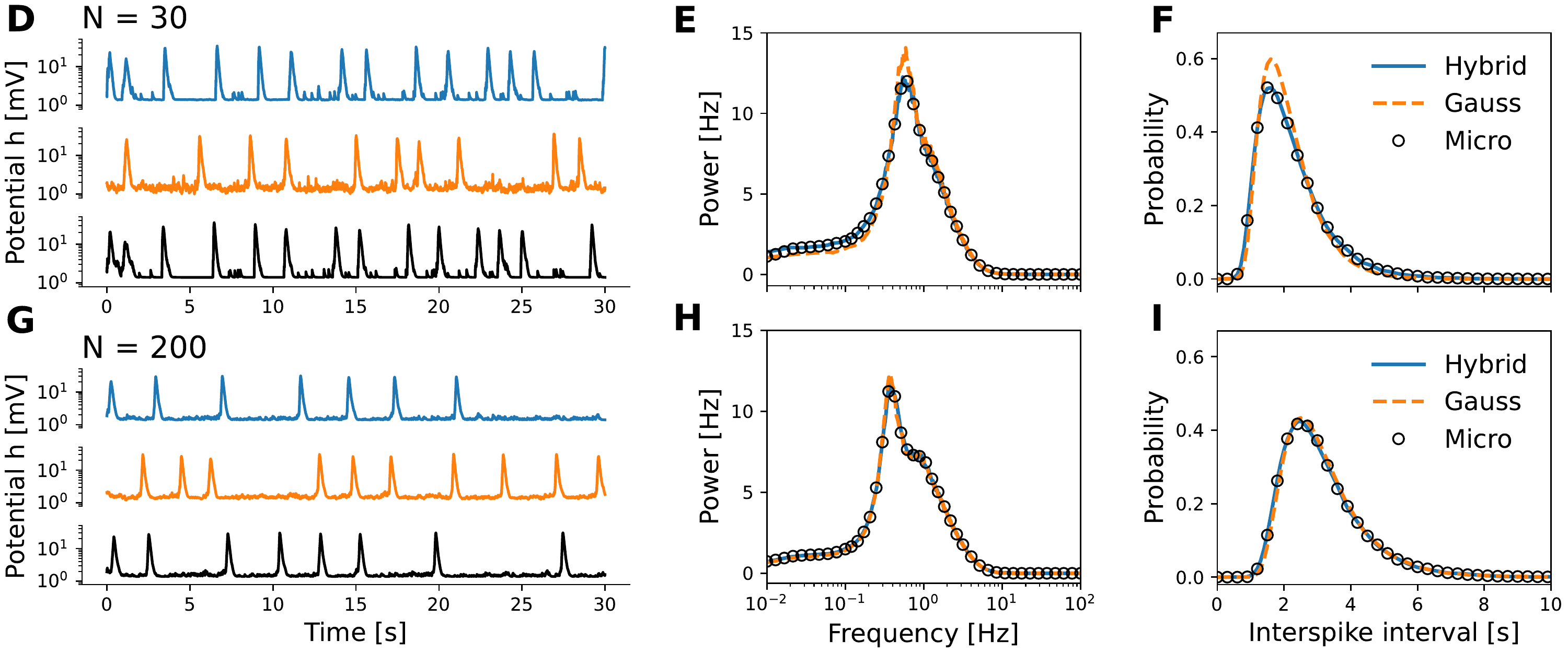}
    \caption{{\bf Population spikes in excitatory populations of finite size.}
    (A) Phase-plane analysis of the macroscopic model (Eq.~\eqref{eq:meso} for $N\to\infty$) reveals the backbone of the metastable dynamics due to the proximity of a separatrix (red-dashed) near the unique stable fixed point (red dot = cross-section of the black-dashed nullclines).
    Trajectories (blue) of the mesoscopic model reproduce population spikes by following the unstable manifold (orange dotted line) of the saddle fixed point (orange diamond).
    Population spikes have variable amplitude and inter-population spike intervals (ISI), see also (B,C).
    (D) The mesoscopic models with hybrid noise (jump-diffusion model; blue) and Gaussian noise (diffusion model; orange) accurately capture finite-size fluctuations in the input potential $h$ -- note the logarithmic y-scale -- and population spikes of the microscopic network dynamics (black) of $N=30$ neurons.
    (E) Power spectra of the input potential $h$ and (F) ISI distributions coincide for all three models. 
    (G-H) same as (D-F) for $N=200$. Statistics are for simulations of length $T_\text{sim} = 100'000$s.}
    \label{fig:meta}
\end{figure}

As a first example, we study the emergence of spontaneous bursts of synchronized activity due to finite-size fluctuations and short-term depression (Fig.~\ref{fig:meta}). To this end, we tune the parameters of our model such that the macroscopic dynamics for $N\to\infty$ exhibits a unique stable fixed point (red dot in Fig.~\ref{fig:meta}A) together with a pair of unstable fixed points (orange, green). In the absence of external inputs or internal finite-size fluctuations, the system will remain in the stable, low-activity state forever. This state, however, is excitable: Fluctuations can lead to rapid, transient excursions of the neural trajectory, when the system is kicked across a separatrix (red-dashed curve = stable manifold emanating from the unstable saddle point (orange diamond)), see the blue traces in Fig.~\ref{fig:meta}A.
During an excursion along the unstable manifold of the saddle point (orange-dotted), the input potential $h(t)$, and with it the population firing rate $f(h)$, rapidly increases, which corresponds to a short synchronized burst of activity. The increased firing of spikes leads to a strong suppression of the depression variable $x$, which in turn pulls the firing rate down. Once the depression variable $x(t)$ has recovered sufficiently, finite-size fluctuations can again trigger a synchronized burst of activity (Fig.~\ref{fig:meta}B,C).

These bursts of activity, called population spikes, have been studied theoretically in the context of STP \cite{TsoUzi00,LoeTso02,GigDec15,SchGer20} and have also been observed experimentally \cite{DewZad06,GigDec15}. Here, we complement the existing literature by pinpointing to the (finite) network size as a possible mechanism for endogenously generated population spikes without the need for external (noisy) inputs. The $N\to\infty$ limit allowed us to draw important insights from the phase-plane analysis of the underlying deterministic structure, which profoundly shapes the mesoscopic network dynamics when considering smaller network sizes $N < \infty$. In principle, the smaller $N$, the larger the fluctuations and the more frequent are the excursions across the separatrix, leading to more population spikes. As can be appreciated in Fig.~\ref{fig:meta}(D,G), our mesoscopic mean-field models -- the diffusion model Eq.\eqref{eq:meso} with Gaussian noise (orange, dashed traces) and the jump-diffusion model Eq.~\eqref{eq:model-depress-3} with hybrid noise (both Gaussian and Poisson; blue traces) -- accurately capture the microscopic network (black traces) both qualitatively and quantitatively. There is a perfect match between the power spectra (Fig.~\ref{fig:meta}E,H) and between the distributions of the inter-population spike intervals (Fig.~\ref{fig:meta}F,I). The slight deviations of the diffusion model for a network of $N=30$ neurons disappear for a network of $N=200$ neurons: as expected, the diffusion approximation becomes better with increased network size. Remarkably, the jump-diffusion model perfectly matches the microscopic network dynamics even when $N=30$.

\subsubsection*{Up-Down dynamics in an excitatory population}
\begin{figure}[!t]
\includegraphics[width=.66\columnwidth]{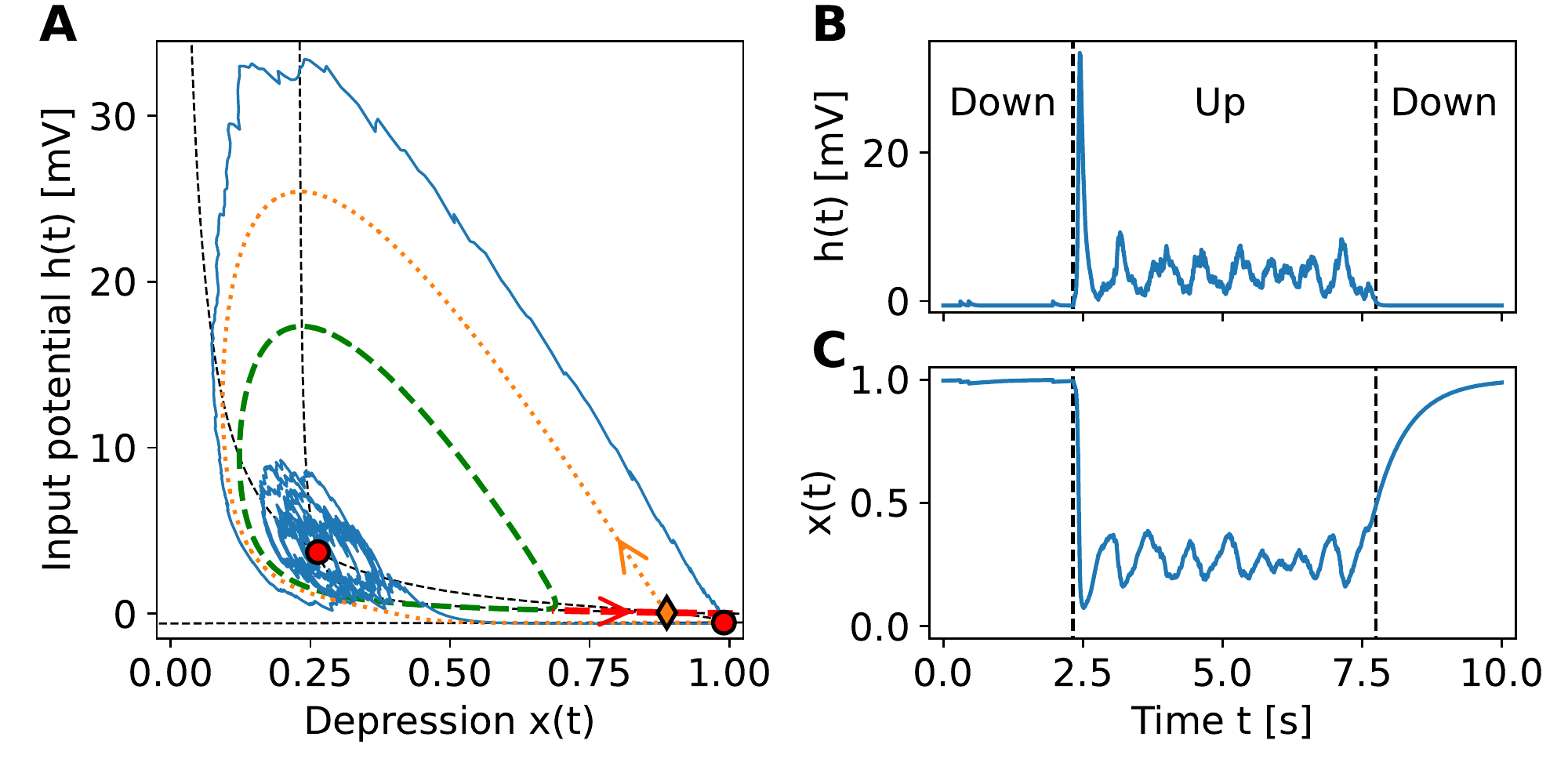}\\ 
\includegraphics[width=\columnwidth]{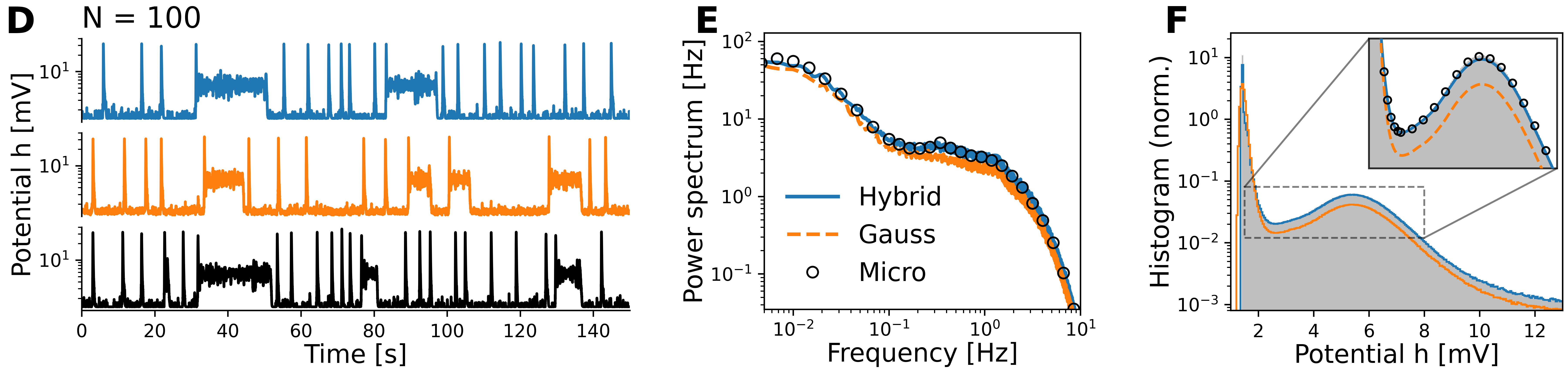}\\[1em]
\caption{{\bf Up-down dynamics due to finite-size fluctuations.} Mesoscopic model reproduces noisy bistable population dynamics. (A) Phase-plane analysis of macroscopic dynamics (Eq.~\eqref{eq:meso} for $N\to\infty$) reveals two stable fixed points (red): a high-activity focus representing the Up, and the low-activity node the Down state of the system. From the saddle fixed point (orange diamond), an unstable (orange dotted line) and a stable manifold (red dashed line) emerge. The latter acts as a separatrix -- trajectories (blue curve) starting from above make an excursion around the unstable limit cycle (green dashed) and converge towards the down state. Finite-size fluctuations can make the trajectory cross the limit cycle into the basin of attraction of the Up state. (B,C) Stochastic trajectory of the mesoscopic dynamics \eqref{eq:meso} with $N=100$ transitioning between Down and Up states. 
(D) The mesoscopic models with hybrid noise (jump-diffusion model; blue) and Gaussian noise (diffusion model; orange) qualitatively capture Up-Down-dynamics of the microscopic network (black).
(E) Power spectrum and (F) histogram of input potential $h$ over simulation of length $T_\text{sim} = 100'000$s.}
\label{fig:updown}
\end{figure}

In a second example, we change the model parameters slightly so that our system now exhibits two co-existing stable fixed points: a high-activity ``Up'' state and a low-activity ``Down'' state. In the macroscopic model, only one of the two states can be realized depending on the initial conditions. 
In the mesoscopic models Eq.~\eqref{eq:meso} and \eqref{eq:model-depress-3}, however, finite-size fluctuations lead to irregular transitions between Up and Down states. 
An exemplary stochastic trajectory in Fig.~\ref{fig:updown}A starts close to the Down state, but soon gets kicked across the separatrix (red-dashed stable manifold of the (orange) saddle fixed point), from where it follows the (orange-dotted) unstable manifold and undergoes a sharp excursion in phase-space, resembling a population spike as described in the foregoing section. 
On its way back to the stable Down state, the trajectory approaches the unstable limit cycle (green dashed) that acts as the boundary of the basin of attraction of the Up state.
Finite-size fluctuations can induce attractor hopping: from the low-activity node (Down state), the trajectory can cross the basin boundary and starts spiraling into the high-activity focus (Up state), until it crosses the basin boundary again and converges towards the low-activity node (Down state), see also Fig.~\ref{fig:updown}(B,C).
The seemingly ongoing oscillations in the Up state are a pure finite-size effect, which will be damped out in the macroscopic model. As an aside, the frequency of the oscillations in the Up state coincides with the imaginary part of the eigenvalue of the high-activity focus, cf.~\cite{DaoDuc15}.

To assess the accuracy of our mesoscopic description of this finite-size induced metastable regime, we performed extensive simulations and compared them to the microscopic network Eq.~\eqref{eq:micro}. In Fig.~\ref{fig:updown}D, we show exemplary time series of the network dynamics for $N=100$ neurons of the jump-diffusion model Eq.~\eqref{eq:model-depress-3} with hybrid noise (blue), of the diffusion model Eq.~\eqref{eq:meso} with Gaussian noise (orange) and of the microscopic model (black). Qualitatively, there is an excellent agreement between micro- and mesoscopic simulations. However, closer inspection of the time series reveal that the Up states in the diffusion model are, on average, of shorter duration than in the microscopic and the jump-diffusion model. This slight shortcoming of the diffusion model also becomes evident when looking at the power spectrum and the bimodal distribution of the input potential $h(t)$ computed  over a long simulation of $T_\text{sim} = 100'000s$ (Fig.~\ref{fig:updown}E and F, respectively). The jump-diffusion model perfectly captures the full statistics of the microscopic network, but the diffusion model slightly underestimates the time spent in the Up states (see the zoom in Fig.~\ref{fig:updown}F), which also manifests in small deviations of the power spectrum. 

To recap, the finite-size induced Up-Down dynamics is very well captured in our mesoscopic description, with excellent agreement between the jump-diffusion model and the microscopic network and only slight deviations of the diffusion model from the true network dynamics for reasonably small network sizes of $N=100$ neurons. As the diffusion approximation requires $N$ to be large, one could expect that the performance of the diffusion model will increase for larger networks. At the same time, however, finite-size fluctuations will become smaller in amplitude, making attractor hopping between Up and Down states more difficult and less frequent. Moreover, as can be seen in  Fig.~\ref{fig:updown}A, the unstable manifold of the saddle fixed point (orange dotted line) leads the neural trajectory close to the basin boundary (green dashed) of the Up state only in a small region of the phase space. In this region, fluctuations need to perturb the neural trajectory in a particular direction so that it can enter (and then also remain within) the basin of attraction of the Up state. In our case, only the jump-diffusion model recapitulates the correct fluctuations, whereas noise in the diffusion model is too diffusive. This discrepancy between Eqs.~\eqref{eq:meso} and \eqref{eq:model-depress-3} can already be anticipated from their $Q$- and $\tilde Q$-dynamics, respectively, which directly influence the finite-size fluctuations. In fact, the steady state profiles of $Q$ and $\tilde Q$ predict that the fluctuations are strongest for intermediate firing rates $f(h)$ and depression levels $0 < x < 1$, that is, right in the aforementioned critical region of the phase space, where a delicate balance between Poisson and Gaussian noise is important to recover the microscopic network dynamics. Consequently, the accuracy of the arguably simpler diffusion model Eq.~\eqref{eq:meso}, and hence its choice over the more complex jump-diffusion model Eq.~\eqref{eq:model-depress-3}, to describe the microscopic network dynamics depends not only on the network size $N$, but also on the dynamical regime under investigation. 

Our study of the Up-Down-dynamics has been guided by the work of Holcman, Tsodyks and co-workers \cite{BarBao05,HolTso06}, who considered a firing rate model with external stochastic input as a necessary ingredient to realize the metastable dynamics and irregular transitions between Up and Down states in a network with STD, which was supported by experimental observations in \cite{DaoDuc15}.
The Up and Down transitions described by our mesoscopic model in Fig.~\ref{fig:updown}(A-C) solely stem from internally generated finite-size fluctuations and, importantly, no external noise is needed.
In addition, the mesoscopic model can explain certain dynamical features that Holcman and Tsodyks ascribed to long-term synaptic plasticity by changing the network size.
Specifically, in \cite{HolTso06} Holcman and Tsodyks attested deviations from typical Up-Down dynamics to external stimulation (by changing the input parameter $\mu$) or to long-term synaptic plasticity (by changing the recurrent coupling strength $J$).
A depolarization injection current (larger $\mu$) had a similar effect as long-term potentiation (LTP; stronger recurrent coupling $J$), which led to longer Up states.
On the other hand, a hyperpolarization injection current (smaller $\mu$), or similarly long-term depression (LTD; smaller $J$), led to shorter Up states and more frequent population spikes.
While our mesoscopic description predicts analogous behavior of the microscopic network when changing $\mu$ and/or $J$, it also predicts that similar effects can be realized by varying the network size $N$. Performing simulations with various network sizes between $N=50$ and $N=150$ while keeping the other parameters unchanged, we found that for $N=100$, metastable activity featuring populations spikes is interspersed with Up states that last on average $10.6$~s (mean taken over all Up states that last at least $1$~s), see Fig.~\ref{fig:updown}D. For smaller networks with $N=50$ (simulations not shown), population spikes become more frequent and Up states become shorter (mean duration $3.1s$): Over a $10'000$~s simulation, Up states followed population spikes in $16.9\%$ of all $2249$ cases for $N=50$, whereas for $N=100$ this only occurred in $15.6\%$ of all $1218$ cases. Larger networks with $N=150$, by contrast, exhibit significantly longer Up states (mean duration $41.9$~s) and even less frequent population spikes (see Fig.~\eqref{fig:supp1} in the Supporting Information). Computing the corresponding histograms of the input potential revealed only for sufficiently large network sizes a second peak at around $5.5$~mV (as in Fig.~\ref{fig:updown}F), which coincides with the Up state in Fig.~\ref{fig:updown}A. Furthermore, the fluctuation-driven oscillations in the long-lasting Up states (Fig.\ref{fig:updown}B,C for $N=100$) become visible in the power spectrum for $N=150$ as another peak at around $1.5$~Hz emerges. This frequency corresponds to the imaginary part of the eigenvalues of the stable focus ($\lambda \approx (-1.54 + 9.24i)$~Hz hence $\mathrm{Im}(\lambda)/(2\pi) \approx 1.5$~Hz).

We can conclude that the network size critically affects the Up-Down dynamics. One may wonder whether dynamically changing the network size by recruiting more or less neurons could be used to dynamically control Up and Down states in biology. One needs to keep in mind, however, that when varying the number $N$ of neurons in our model, we have simultaneously rescaled the synaptic weights in proportion to $1/N$. Although this specificic change of parameters is useful for the theoretical analysis, a corresponding biological implementation is difficult to perceive.
In any case, for any given network size and synaptic weights, our mesoscopic mean-field models Eq.~\eqref{eq:meso} and \eqref{eq:model-depress-3} provide accurate descriptions of the microscopic network with short-term synaptic depression and therefore allow for a systematic analysis of how finite-size fluctuations contribute to and shape Up- and Down-states dynamics. Thus, in principle, analyzing the dependence on network size with fixed synaptic weights also seems to be feasible.





\subsection*{Mesoscopic model for hippocampal replays}
We now turn to a more complex biological example for metastability in neural circuits: the spontaneous replay of activity sequences across hippocampal place cells  \cite{Buz15,Fos17}. Sequential activation patterns of place cells have been widely observed in experiments when an animal explores its environment \cite{OkeDos71,Oke76,HarCol09,MabAck18} and have been related to  neural representations of animal trajectories. Distinct sequences of place cell activation show up when entering novel environments. Hence, sequential activation patterns during exploration provide unique signatures for each environment and may subserve navigation and spatial learning \cite{KniKud95,MosKro08,HarLev14,TheRov18}. Presumably, the animal forms an internal representation, or map, of the corresponding environment during exploration through such sequential activation \cite{WuFos14} that can later be replayed spontaneously, i.e. in the absence of sensory input---a feature that is believed to contribute to memory consolidation and retrieval \cite{DerMos10,DupOne10,Pfe20} as well as to route planning \cite{PfeFos13,OlaBus18}.
The spontaneous replay occurs within burst-like sharp-wave/ripples (SWRs) during quiet wakefulness \cite{FosWil06,KarFra09,CarJad11} and sleep \cite{WilMcN94,LeeWil02}, typically has a much faster, compressed time scale \cite{DavKlo09,FosWil06,LeeWil02} and the replayed trajectories can either be in the originally experienced order or backwards \cite{DraTon11}. Spontaneous replay events appear and disappear abruptly and repeatedly, and can therefore be regarded as metastable states separated by states of low activity. 

In previous approaches to model metastable sequential activation patterns, such as in hippocampal replay, it has been difficult to accommodate both the capability to endogenously generate sequential activity \cite{SomKan86,Kle86,PerBru20} and trial-to-trial temporal variability \cite{LitDoi12,MazFon15}. Successful candidate mechanisms to account for both characteristics have been implemented with the help of firing rate models. Recanatesi et al. \cite{RecPer22} proposed a two-area mesoscale attractor network very much in the spirit of the winnerless competition model by Seliger, Tsimring and Rabinovich \cite{SelTsi03}, in which the combination of asymmetric synaptic connectivity, arising from reciprocal coupling between fast and slow systems, with stochastic synaptic efficacy is crucial for the generation of sequences. 
Alternatively, Romani and Tsodyks \cite{RomTso15} proposed a fully deterministic mechanism for hippocampal replay based on short-term depression (STD) and without the need for asymmetric connectivity, see also \cite{TheRov18}: A ring-attractor network model with symmetric synaptic connectivity with local excitation and long-range inhibition exhibits multistability between a global quiescent state and various spatially localized bump states \cite{Ama77,BenBar95,BurFie12}. STD, as a slow fatigue mechanism, destabilizes these bumps and gives rise to traveling wave states\cite{YorRos09}. In combination with the high-dimensional character of the network, the resulting dynamics appears effectively stochastic and spontaneously switches between quiescence and metastable traveling waves.
Both the approaches by Recanatesi et al. \cite{RecPer22} and by Romani and Tsodyks \cite{RomTso15} rely on heuristic firing-rate models, and as such both of them suffer from the limitations already discussed in the Introduction. In particular, neither can account for a systematic investigation of finite-size effects on metastable activity. 
In addition, it is unclear to what extent the deterministic Romani-Tsodyks model describes neural  variability. For these reasons, we use our mesoscopic theory to construct a stochastic ring-attractor network model from a microscale model, and thereby provide an alternative mechanistic description of hippocampal replay with a direct link to microscopic networks of spiking neurons.

\subsubsection*{Microscopic and mesoscopic multi-population model of place cells}

We aim for a mesoscopic description of place cells in area CA3 of the hippocampus.
Following Romani and Tsodyks \cite{RomTso15}, we consider a network of neuronal populations, where each population is a group of neurons with highly overlapping place fields. We assume that the full map of the environment is covered by in total $M$ populations each containing $N$ neurons.  
The activity of individual neurons $j \in \{1,\dots,N\} $ of a given population $\alpha \in \{1,\dots,M\}$ are described by the spike trains $s_j^\alpha(t) = \sum_k \delta(t-t_{j,k}^\alpha)$ associated with the spike times $\{t_{j,k}^\alpha\}$.  They are modeled as stochastic point processes with intensity $r^\alpha(t) = f\big( h^\alpha(t^-) \big)$, where the input potentials $h^\alpha$ (identical for all neurons in population $\alpha$) are given by the following neuronal dynamics with STD:
\begin{subequations}\label{eq:ring_micro}
\begin{align}
    \od{h^\alpha}{t}&=\frac{\mu^\alpha-h^\alpha}{\tau}+\frac{1}{M}\sum_{\beta=1}^M \frac{J_{\alpha\beta}}{N}\sum_{j=1}^N U_0x_j^\beta(t^-)s_j^\beta(t),\label{eq:ring_micro_a}\\
    \od{x_j^\alpha}{t}&=\frac{1-x_j^\alpha}{\tau_D}-U_0x_j^\alpha(t^-)s_j^\alpha(t).
\end{align}
\end{subequations}
Here, the input potential $h^\alpha(t)$ integrates the external input $\mu^\alpha$ (common to all neurons in population $\alpha$) and the recurrent input. The latter consists of contributions from the neurons in the same population but also from all the other populations $\beta\neq \alpha$, weighted with a synaptic strength $J_{\alpha\beta}$ that depends on the distance between the place fields of the corresponding populations (as detailed below).
The resulting recurrent connectivity of the network with weights $J_{\alpha\beta}$ is assumed to encode the internal representation of one (or multiple) environment(s) that the animal has explored recently.
It should be noted that the recurrent weights of an internal map can also be ``learnt'' via spike-timing-dependent synaptic plasticity (STDP) during active exploration of the environment, see, e.g., \cite{TheRov18,EckBag22}. Here, however, we assume for simplicity that the animal has already internalized the relevant environments and that the corresponding internal maps are hardwired (at least on the relevant time scale) within the synaptic connectivity matrix $\{J_{\alpha\beta}\}_{\alpha,\beta}$ in a Hopfield-like manner \cite{Hop82}.

Analogously to the Results subsection ``Diffusion model for mesoscopic dynamics'', we can reduce the microscopic dynamics Eq.~\eqref{eq:ring_micro} to a mesoscopic mean-field model. 
As before, we introduce the mesoscopic quantities $x^\alpha(t)$ and $Q^\alpha(t)$ that correspond to the first and second moment, respectively, of the depression variables $x_i^\alpha(t)$ for population $\alpha \in \{1,\dots,M\}$.
We then obtain the mesoscopic dynamics
\begin{subequations}\label{eq:ring_meso}
\begin{align}
    \od{h^\alpha}{t}&=\frac{\mu^\alpha-h^\alpha}{\tau}+\frac{1}{M}\sum_{\beta=1}^M  J_{\alpha\beta}U_0 \Big[ x^\beta f\big(h^\beta\big) + \sqrt{\frac{Q^\beta f(h^\beta)}{N}} \xi^\beta(t) \Big],\label{eq:ring_meso_h}\\
    \od{x^\alpha}{t}&=\frac{1-x^\alpha}{\tau_D}-U_0\Big[ x^\alpha f\big(h^\alpha\big) + \sqrt{\frac{Q^\alpha f(h^\alpha)}{N}} \xi^\alpha(t) \Big],\\
    \od{Q^\alpha}{t}&=2\frac{x^\alpha-Q^\alpha}{\tau_D}-U_0(2-U_0)Q^\alpha f(h^\alpha),
\end{align}
\end{subequations}
with Gaussian white noises $\xi^\alpha(t)$ obeying $\langle \xi^\alpha(t) \rangle = 0$ and $\langle \xi^\alpha(t)\xi^\beta(s) \rangle = \delta_{\alpha,\beta}\delta(t-s)$ for all $\alpha,\beta \in \{1,\dots,M\}$.
We only present the mesoscopic ring-attractor network Eq.~\eqref{eq:ring_meso} under the diffusion approximation, but remark that it can readily be extended to a jump-diffusion model as in Eq.~\eqref{eq:model-depress-3}. 
In the following, we use the simpler diffusion model because it already faithfully reproduces microscopic simulations and is sufficient to study the mechanisms of some recent experimental observations. For one, there is a need for stochastic network models of hippocampal replay as replay patterns resemble Brownian diffusion \cite{SteBar19} or even super-diffusion \cite{KraDru22}, and interburst intervals exhibit significant variability. Moreover, replay episodes do not always draw a smooth continuous path, but often follow a jumpy, discontinuous trajectory \cite{PfeFos15,DenGil21}. 

\subsubsection*{A circular environment}

\begin{figure}[p]
  \begin{adjustwidth}{-1.0in}{-1in}
    \centering{
\includegraphics[width=7.5in]{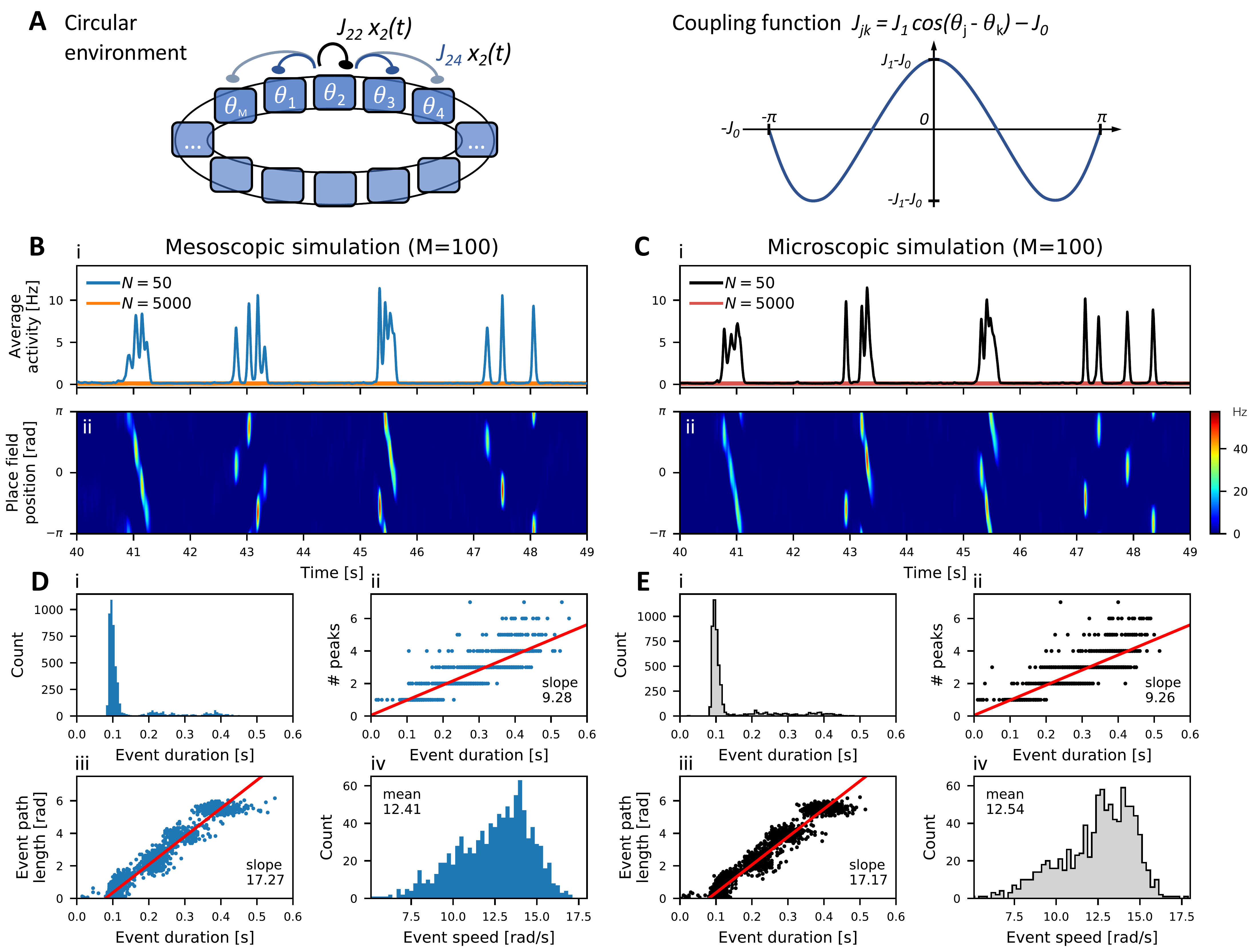}
}
\end{adjustwidth}
\captionsetup{width=7.5in}
\caption{{\bf Hippocampal replay in micro- and mesoscopic ring-attractor network model.}
(A) Ring-attractor model of $M$ population units of $N$ LNP spiking neurons with STD. Synaptic weights $J_{\alpha,\beta}$ are excitatory for units with nearby place field positions $\theta_\alpha$ and inhibitory at longer distances, see the coupling function on the right. 
(B) Mesoscopic and (C) microscopic network simulations reveal (i) spontaneous bursts of the averaged activity, resembling SWRs, during which replay patterns evolve as (ii) metastable traveling waves, or nonlocal replay events (NLE), along the circular environment---the expected activity $r_j = f(h_j)$ at location $\theta_j=2\pi j/M$, $j=1,\dots,M$ is color-coded.
Statistics of the (D) mesoscopic and (E) microscopic simulations perfectly match each other with respect to: (i) the distribution of event duration, (ii) the correlation between the number of peaks per burst and its duration, (iii) the correlation between the length of the traveled path during an event and its duration, as well as (iv) the distribution of average bump speed during an event, computed from events with more than one peak; see Methods for more details.
}
\label{fig:1Env}
\end{figure}

To begin with, we consider a single circular environment and assume that, after exploration, the animal has learnt an internal representation (map) of the corresponding environment, which is encoded in the synaptic connectivity of the place cells in hippocampal area CA3. Accordingly, we assign to all neurons in population $\alpha\in\{1,\dots,M\}$ 
a place field at angle $\theta_\alpha = 2\pi \alpha/M$, so that the place field locations are equally spaced on a ring, see Fig.~\ref{fig:1Env}A.
The synaptic strength $J_{\alpha\beta}$ depends on the distance between locations according to
\begin{equation}
    J_{\alpha\beta} = J_1 \cos( \theta_\alpha-\theta_\beta ) - J_0 = J_1 \cos( 2\pi (\alpha-\beta)/M ) - J_0,
\end{equation}
where $J_1$ scales the strength of map-specific interactions and $J_0$ corresponds to uniform feedback inhibition. 
For $J_1 > J_0 \ge 0$, populations with adjacent place fields excite each other, whereas populations with place fields far apart from each other are inhibitory.
This form of symmetric interaction is known to generate spatially coherent activity, leading to so-called bump attractors \cite{Ama77,BenBar95,TsoSej95,HanSom98}.
STD has been shown to destabilize stationary bumps that move around the environment \cite{YorRos09}, creating burst-like nonlocal traveling wave events (NLEs) that resemble hippocampal replay patterns on a population level.

In our network simulations, we consider $M=100$ populations around the ring with $N=50$ neurons each. The external input is homogeneous, i.e. $\mu^\alpha=\mu$ for all $\alpha=1,\dotsc,M$. As shown in Fig.~\ref{fig:1Env}Bi, the network spontaneously generates a burst of elevated activity with one or more peaks, that lasts up to a few hundred milliseconds and is then terminated due to STD. Another burst is generated only after a recovery period--the so-called interburst interval (IBI).
The intermittently elevated activity on the ring is strongly localized because of the synaptic connectivity. STD makes the localized bump move to neighboring place field locations and, thus, gives rise to an NLE, Fig.~\ref{fig:1Env}Bii; we define an NLE (=nonlocal replay event) as a burst of the averaged activity with more than one peak, see also Methods.
As the elevated activity locally alters the spatial profile of the slow synaptic depression variable, highly irregular activity patterns emerge with bursts that start at seemingly random locations, travel in either backward or forward direction, and vary both in duration and distance (note, however, that a single wave of activity never travels further than once around the circle). 
This type of metastable dynamics strongly resembles the behavior observed by Romani and Tsodyks in their deterministic firing rate model \cite{RomTso15}, hereafter referred to as the RT model, but with the important difference that here the emergence of metastability is a finite-size effect---increasing the population size from $N=50$ to $N=5000$ renders any initiation of burst-like activity impossible (see the orange curve in Fig.~\ref{fig:1Env}Bi). Put differently, our model reveals a novel dynamical regime, in which metastable burst states are fluctuation driven, in contrast to the RT model, where bursts are induced deterministically when depression slowly abates.

Comparing the simulations of our mesoscopic model with the microscopic network, we find an excellent agreement both from a qualitative (Fig.~\ref{fig:1Env}B,C) and from a quantitative perspective (Fig.~\ref{fig:1Env}D,E), where we follow the statistical analysis of \cite{RomTso15}, see also Methods.
 For comparison, we considered two deterministic models with fatigue-induced bursts: (i) the original RT model \cite{RomTso15}, and (ii) our model with $N\to\infty$ and slightly increased external drive ($\mu=-0.9$ instead of $\mu=-1$). This second model, which will be referred to as the \emph{macroscopic} model in the following, essentially reproduces the dynamics of the original RT model \cite{RomTso15}. We included this second model in the comparison  (see also Fig.~\ref{fig:supp2} in the Supporting Information) because the macroscopic limit $N\to\infty$ of our mesoscopic model yields slightly different model equations compared to the heuristic RT model.
A closer inspection of the statistical properties of NLEs and IBIs reveals that mesoscopic and microscopic simulations not only match almost perfectly, but the fluctuation-induced bursts may also have more biologically realistic properties than the fatigue-induced regime.
For example, experiments in rodents exposed to long linear tracks \cite{DavKlo09,GupvdM10} had an experimentally estimated number of ~$10$ SWRs/s. Assuming that each peak in the average activity corresponds to a SWR, the micro- and mesoscopic models closely match the experimental observations with $9.3$ SWRs/s in contrast to less than $8$ SWRs/s in the macroscopic and original RT model, see Methods for more details. 

\begin{figure}[t]
\begin{adjustwidth}{-1.0in}{-1in}
\centering{
\includegraphics[width=7.5in]{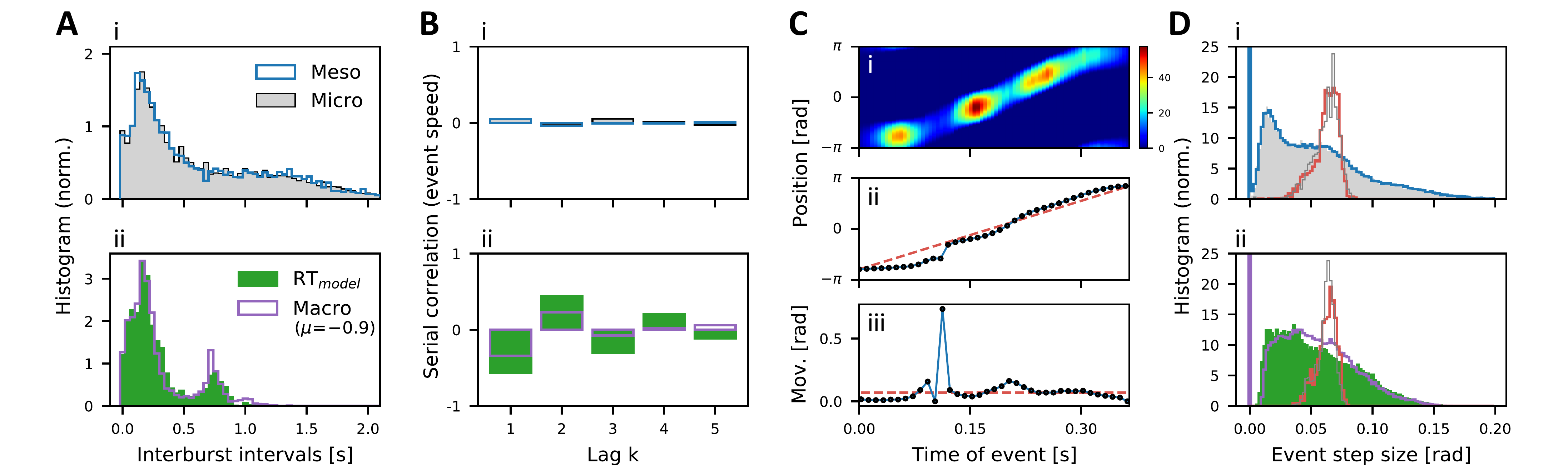}
}
\end{adjustwidth}
\captionsetup{width=7.5in}
\caption{{\bf Comparison of fluctuation-induced and depression induced hippocampal replay dynamics.}
(A) Interburst-interval distributions and (B) serial correlations of the event speeds of consecutive NLEs for (i) the meso- and microscopic models and (ii) the deterministic models (green: original Romani-Tsodyks model \cite{RomTso15}, purple: macroscopic model in the fatigue-driven regime obtained by setting $N^\alpha\to\infty$ and $\mu=-0.9$ (instead of $\mu=-1$)).
(C) On shorter time scales, all models capture the discontinuous nature of the replayed trajectory: (i) Single NLE of the mesoscopic model. (ii) Place field positions decoded using PVA in bins of 50ms (black dots). This "replayed trajectory" deviates from a straight line (red-dashed) corresponding to a hypothetical uniform motion. (iii) Increments of the movement trajectory exhibit strongly irregular movement features (black dots) in contrast to the constant increments expected for a straight line (red-dashed).
(D) (i): The distributions of reconstructed step-sizes of the meso- (blue) and microscopic models (gray histogram) coincide and strongly deviate from the narrow distributions of average step sizes (computed for each NLE by fitting straight lines to individual movement trajectories) for the meso- (red) and microscopic models (gray thin line). The average step sizes vary for different NLE's, causing the non-zero width of their distribution. (ii) Similar behavior is observed for the macroscopic model (purple; with increased external input $\mu = -1.0 \mapsto -0.9$) and the Romani-Tsodyks (RT) model (green histogram); red/gray thin lines correspond to average step sizes in the macro/RT-models, respectively. See main text and Methods for more details.
}
\label{fig:1EnvB}
\end{figure}

Furthermore, our fluctuation-driven model reveals larger temporal variability with a unimodal IBI distribution (Fig.~\ref{fig:1EnvB}A) and low serial correlations of the event speeds (Fig.~\ref{fig:1EnvB}B). 
In marked contrast, in the macroscopic and RT models of fatigue-induced bursts, the IBI distributions are bimodal, and clearly not exponentially distributed as observed experimentally \cite{AxmElg08,SchKal14,Buz15} and expected for a Poisson process. By contrast, our fluctuation-driven model shows the tendency towards exponential IBI distributions with longer tails, showing in particular a larger mean ($0.65$~s vs. $0.29$~s) and a higher coefficient of variation (CV $= 0.85$ vs. $0.79$).
Moreover, the macroscopic and RT models exhibit strong correlations between forward and backward replay events as seen in the serial correlations of the event speed (Fig.~\ref{fig:1EnvB}Bii). The alternating structure of the serial correlation coefficient with strong anti-correlations at lag $1$ means that forward and backward motion alternate almost perfectly. In contrast, the motion directions in the sequence of NLEs in our fluctuation-driven model are almost uncorrelated (Fig.~\ref{fig:1EnvB}Bi). 
Another difference to the deterministic models is that the onset location of the fluctuation-induced NLEs is independent of the offset location of the preceding event. By contrast, in the deterministic fatigue-induced (macroscopic and RT) models, the activity bursts start at the location where the slow depression variable has had most time to recover, leading to more regular event patterns (see also \cite{RomTso15} and Fig.~\ref{fig:supp2}).

On shorter time scales, all models (micro-, meso-, macroscopic and RT) exhibit the experimentally observed discontinuous nature of replay events \cite{PfeFos15}. In Fig.~\ref{fig:1EnvB}Ci, we zoom into one exemplary NLE of around $350$ms, during which a metastable wave travels around the ring in backward direction. The place field activity (color coded) varies critically in location and activity. Binning activity per location in moving $50$ms windows, we estimate the animal's (hypothetical) position along the ring by applying the population vector average (PVA; \cite{KimRou17}), see Fig.~\ref{fig:1EnvB}Cii. The decoded trajectory (black dots) does not follow a smooth path, but the step sizes vary irregularly in length (Fig.~\ref{fig:1EnvB}Ciii). In Fig.~\ref{fig:1EnvB}D, we show the step-size distributions for the four different models, featuring a large number of very short steps and a long tail of larger steps. The broad distributions significantly differ from the narrow step-size distribution that would be expected if the movement trajectory was uniform (thin lines in Fig.~\ref{fig:1EnvB}D). More precisely, the latter distribution describes the variation of the average step sizes (computed for each NLE) across different NLE's, which corresponds to approximating each replay trajectory by a straight line (red-dashed curves in Fig.~\ref{fig:1EnvB}C), see also \cite{PfeFos15} for more details on computing the step-size distributions.

In conclusion, the mesoscopic model Eq.~\eqref{eq:ring_meso} perfectly recovers the metastable replay dynamics of the microscopic network model Eq.~\eqref{eq:ring_micro}. While our fluctuation-driven model is similar to the deterministic models with respect to the structure of single replay trajectories (Fig.~\ref{fig:1EnvB}C,D), the fluctuation-driven model features unimodal IBI distributions and low serial correlations of motion directions, in marked contrast to the bimodal IBI distribution and the strong serial correlations of the deterministic fatigue-induced replay model. Thus, the IBI distribution and the sequence of motion directions provide experimentally testable predictions that may be useful to disentangle the contributions of deterministic and stochastic sources of metastable hippocampal dynamics.


\subsubsection*{Multiple circular environments}
In a next step, we assume that an animal has internalized multiple environments. The ability to code for spatial locations in multiple environments is considered one of the hallmarks of place cell activity in the hippocampus.
Experimental results have shown that rodents, when exposed to two distinct environments of similar shape, most place cells are active in only one environment. However, a few place cells are active in both environments but they typically exhibit place fields at different spatial locations, which is referred to as global remapping \cite{MulKub87,BosMul91}. Replay events can then be observed in both neural maps corresponding to each of the two environments \cite{KarFra09}, see also \cite{DraTon13,GriSch20}. 

Following \cite{RomTso15}, we consider $K$ circular environments and store their respective maps within the synaptic connectivity $J_{\alpha,\beta}$ of the network model Eq.~\eqref{eq:ring_micro}. To this end, we endow each population $\alpha$ with a binary vector of selectivities for these $K$ environments, $\zeta_\alpha = (\zeta_\alpha^1,\dots,\zeta_\alpha^K) \in \{0,1\}^K$, where $\zeta_\alpha^k = 1$ indicates that the neurons in population $\alpha$ are selective for environment $k \in \{1,\dots,K\}$ (i.e., the neurons contribute to the encoding of this environment) \cite{RomTso15,SolYou14}. 
Otherwise, if $\zeta_\alpha^k=0$,  population $\alpha$ is not selective for environment $k$.
Selectivity to particular environments is assigned randomly, but with the constraint that $\sum_{\alpha=1}^M \zeta_\alpha^k = f M$ with $f \in [0,1]$, i.e. exactly $f M$ populations are selective for environment $k$.
Furthermore, we introduce place field locations $\theta_{\tilde\alpha}^k = 2\pi \tilde\alpha/(fM)$ for each environment $k$ and randomly assign a unique place field angle to each of the $fM$ populations $\tilde\alpha \in \{1,\dots,fM\}$ selective for environment $k$.
We then define the synaptic weights as
\begin{equation}\label{eq:J_3env}
    J_{\alpha\beta} = \frac{1}{f} \sum_{\mu=1}^K J_1 \zeta_\alpha^k \zeta_\beta^k \cos( \theta_\alpha^k - \theta_\beta^k) - J_0,
\end{equation}
where map-specific interactions of strength $J_1$ only occur within environments, and $J_0$ represents uniform feedback inhibition as before \cite{RomTso15}. 
In our simulations, we use $K=3$, $f=0.3$ and $M=300$, and find parameters of synaptic strength $J_0$ and $J_1$ as well as the homogeneous external input $\mu$ such that replay events are again a pure finite-size effect: When increasing the population size from $N=50$ to $N=500$, the network remains in a quiescent state and bursts of elevated activity no longer occur in our simulation.

\begin{figure}[p]
\begin{adjustwidth}{-1.0in}{-1in}
\centering
\includegraphics[width=7.25in]{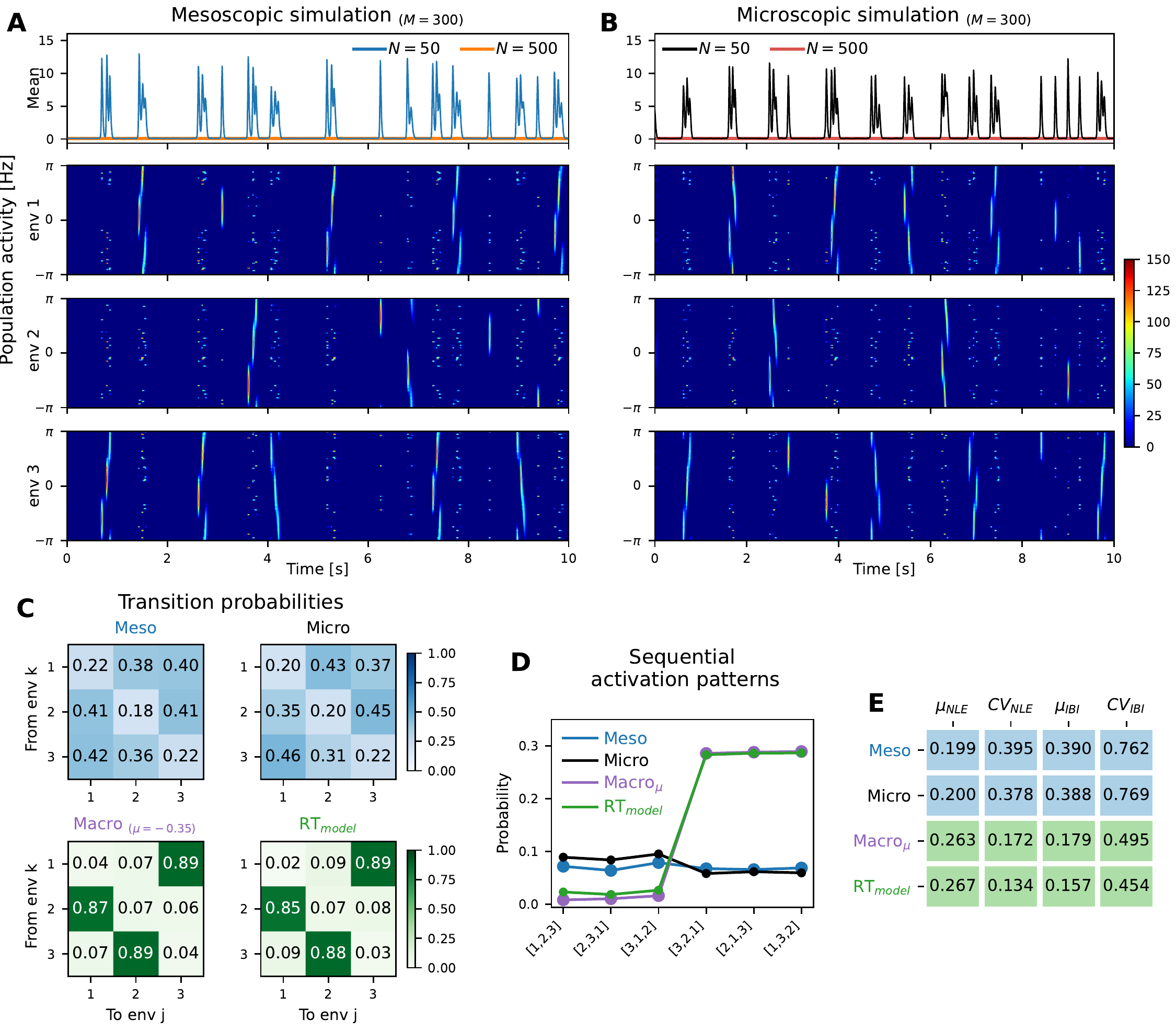}
\end{adjustwidth}
\captionsetup{width=7.5in}
\caption{{\bf Spontaneous replay switches between multiple environments.}
In the (A) mesoscopic and (B) microscopic ring-attractor network storing multiple environments, metastable replay dynamics spontaneously emerge due to finite-size fluctuations when decreasing the population size from $N=500$ (orange/red in panels i) to $N=50$ (blue/black) per unit.
Nonlocal replay events (NLEs) occur randomly in exclusively one of three environments, while activity in the respective other two is suppressed.
The resulting activation sequences of replayed environments--in (A) the activation sequence reads 313123122312321--are analyzed with respect to (C) the transition probabilities between subsequently active environments and (D) sequential activation patterns. 
In the meso- and microscopic models, transitions from environment $k$ to $j$ are equally likely for all pairs $(k,j) \in \{1,2,3\}^2$. But the deterministic (macro and RT) models show a clear preference for transitions $1\to3\to2\to1$, which is also apparent in the high probability of the corresponding subsequences of three distinct, subsequently active environments.
(E) Larger heterogeneity with respect to NLE duration and interburst intervals (IBI), as assessed by the respective means $\mu$ and coefficients of variation $CV$, further distinguishes the more variable metastable regime of the micro-/mesoscopic vis-\'a-vis macroscopic/deterministic models.
}
\label{fig:3env}
\end{figure}

As can be seen in Fig.~\ref{fig:3env}, bursts of elevated activity occur spontaneously and with high temporal variability. During these nonlocal replay events (NLE) a metastable traveling wave state is generated randomly in one of the three stored environments; activity in the other environments is suppressed due to global inhibition. As expected, our meso- and microscopic network simulations show qualitatively very similar behavior (cf.~Fig.~\ref{fig:3env}A and B). The metastable dynamics exhibit high variability with respect to the duration of NLEs and the interburst intervals (Fig.~\ref{fig:3env}E), the length of the traveled path during a NLE within the active environment, as well as the order of environment activation.

In more detail, we statistically analyzed the patterns of sequential activations. For instance, in Fig.~\ref{fig:3env}A the order of environment activation reads 313123122312321. To quantify whether replay of distinct environments, i.e.\ their order of activation, is random or correlated, we first computed the transition probabilities between environments. While the stochastic (meso- and microscopic) models did not show strong preference for any environment transition, in the deterministic (macroscopic and RT) models, preferred transitions were clearly visible (Fig.~\ref{fig:3env}C).
Moreover, we checked whether particular sequences of recalled environments are more probable than others, which may point at some (hidden) deterministic origin of sequence generation and recall. In order to avoid spurious deterministic effects that may be inherited from asymmetries in the synaptic weights, we constructed the selectivity vectors $\zeta_\alpha$ symmetrically and guaranteed that bursts within the environments were equally distributed, see Table~\ref{table:3env} in the Methods section. The deterministic (macroscopic and RT) models, nonetheless, exhibited a strong preference for specific order of environment activation ($3\to2\to1\to3$ in the example on Fig.~\ref{fig:3env}). By contrast, the stochastic models did not show any preference for a particular order, Fig.~\ref{fig:3env}D, which underlines once more the strong variability in the finite-size induced metastable dynamical regime of hippocampal replay.

\section*{Discussion}

To better understand the mechanisms of emerging collective dynamics and metastability in neural networks, low-dimensional mean-field models have become indispensable in theoretical, computational and systems neuroscience. 
In this paper, we have proposed a novel mesoscopic mean-field model for networks of spiking neurons with short-term synaptic plasticity. This mesoscopic model readily allows for systematically analyzing the effect of finite-size fluctuations on metastable dynamics.
Following a bottom-up approach, we have derived simple stochastic differential equations for networks consisting of a finite number of Linear-Nonlinear Poisson (LNP) spiking neurons with dynamic synapses undergoing short-term synaptic plasticity (STP). The mesoscopic model comes in two variants: First, a jump-diffusion model captures the network dynamics of only a few neurons with high accuracy thanks to a hybrid formulation of the finite-size-induced fluctuations, which takes the shot-noise properties of the spike-train inputs into account. 
Second, using a diffusion approximation, we obtained an even simpler diffusion model for pure short-term depression, whose accuracy naturally improves with increasing network size. Noteworthy, its accuracy also depends on the dynamical regime under investigation, e.g., when the skewness of the shot noise critically affects the transitions to the Up-states. Nonetheless, as we showed above, the mesoscopic diffusion model is able to capture the microscopic network dynamics formidably well, allowing us to uncover finite-size induced population spikes, spontaneous transitions between Up and Down states, and a novel dynamical regime of quasi-traveling waves as a putative mechanism for fluctuation-driven hippocampal replay. 

\subsubsection*{Modeling population spikes and Up-Down dynamics}
In the modeling literature, theoretical models of metastability are typically based on an interplay between the network's tendency to posit itself in a self-excitable dynamical regime and some fatigue mechanism that generates activity-dependent self-inhibition in response to elevated network activity \cite{MilMih10,LevHer07}. Such a fatigue mechanism can be implemented in neural networks via neural spike-frequency adaptation (SFA) or via synaptic short-term depression (STD). Here, we focused on STD, but acknowledge that similar behavior can, in principle, also be achieved with SFA. 
One possibility of self-excitability is that a stable low-activity state of asynchronous activity is close to a Hopf bifurcation, at which it becomes destabilized in favor of stable global oscillations. In the subcritical regime, noise can promote transient departures from the fixed point, resembling populations spikes \cite{GigDec15}. Another possibility is that the system exhibits two stable fixed points, a high-activity (Up) and a low-activity (Down) state. Switching between the states can be induced by internal or external noise. In addition, the Up state can get destabilized in a saddle-node bifurcation by an adaptive fatigue mechanism described above \cite{GigMat07,MejKap10,JerRox17,LevBuz19}. These scenarios are effectively low-dimensional and, in consequence, firing rate models describing the mean population activity 
can successfully explain metastable dynamics. As noted in the Introduction, however, the firing rate models largely miss a clear link to microscopic circuit models. While neuronal and synaptic properties can be partly accounted for by mean-field models, incorporating biologically realistic fluctuations in such models is often doomed to failure as it lacks a rigorous footing. Neglecting fluctuations may explain purely deterministic, fatigue-induced metastable activity patterns with low variability. Experimentally observed metastability in the brain, however, shows larger variability suggesting fluctuation-induced metastable dynamics. Fluctuations can have manifold origins that range from external noisy (cortical or thalamic) inputs or background noise \cite{RomAmi06}, via specific network connectivity topologies (random, sparse, or clustered) \cite{EckJac08,SheVol08,MarTso12,LucBen14,PirRic15}, heterogeneity \cite{diSVil18}, (loose or strong) balance between excitation and inhibition, up to finite-size effects \cite{SouCho07,BenCow10,GigDec15}, or even a combination of them \cite{TouHer12,diVRom19}. On the mesoscopic scale, such fluctuations are often modeled heuristically by adding noise terms to the mean-field equations \emph{ad hoc}.

With the mesoscopic mean-field model that we have proposed here, we restricted ourselves to explaining the fluctuations observed on a network level that are due to a finite number of neurons. To minimize confounding factors, we considered just one excitatory population of Poisson spiking neurons, all-to-all coupling, and no external noise.
Indeed, it has been shown that inhibition is not necessarily needed to generate population spikes and Up-Down transitions, neither is short-term synaptic facilitation needed; see \cite{BarBao05,HolTso06,DaoDuc15} who used a mean-field model with STD and Gaussian white noise in the voltage dynamics. Building on their previous work and complementing that mean-field model by an additional facilitation dynamics, Holcman and co-workers more recently provided an improved, and analytically tractable, description of network bursts, Up-Down dynamics and slow oscillations that is compatible with experimental recordings on various scales \cite{DaoLee15,ZonHol21} and allows for detailed stochastic analysis \cite{ZonHol21Com,ZonHol21PRR,ZonHol22}.
In the Methods, we extend our mesoscopic description considering also short-term facilitation. Our mesoscopic mean-field model Eq.~\eqref{eq:meso} yields exactly the same deterministic dynamics of their depression(-facilitation) mean-field model in the limit $N\to\infty$, see also \cite{BarTso07}.
The important difference, however, is how to deal with noise. While Holcman, Tsodyks and co-workers rather vaguely motivated an additive Gaussian white noise term that is meant to represent the fluctuations from independent vesicular release events and/or closings and openings of voltage gated channels, we here provide an explicit and rigorous derivation of the multiplicative noise terms in our mesoscopic mean-field model. Our model can thus accurately account for the finite-size fluctuations of the microscopic network including the thereby induced heterogeneity of synaptic depression across the neurons.

Previous approaches to model finite-size effects either included a multiplicative noise term ad hoc in the firing rate equation \cite{BruHak99,SpiGer99} or dwelled on a master equation formalism leading to coarse-grained phenomenological models of collective activity dynamics \cite{BuiCow07,ElBDes09,Bre10,diSVil18}. By contrast, we here derived Langevin equations for the mesoscopic population dynamics directly from the underlying microscopic network of finitely many Poisson spiking neurons. In our bottom-up approach we explicitly take into account fluctuations due to the variability of the individual depression variables across synapses, that are typically neglected in the other approaches. Our resulting nonlinear mesoscopic model remains nonetheless simple enough---thanks to a minimal microscopic network model considered here, but see below for possible extensions towards more biological realism---to allow for a systematic analysis of finite-size induced metastable dynamics in the presence of a slow fatigue mechanism in form of STD.


\subsubsection*{Predictions and possible functional roles of variable hippocampal dynamics}

Given the recent success in explaining a wide range of experimental observations on hippocampal dynamics by means of firing-rate models with STD \cite{RomTso15,TheRov18}, we extended our mesoscopic description for a single population to a ring-attractor model of spiking neurons. Aiming at a minimal spiking neuron network model that can offer unique insights in the generative mechanisms of hippocampal dynamics in area CA3---in particular those underlying sharp waves and bidirectional activity replay---, we ignored various degrees of biological plausibility on purpose, but see below for possible extensions. Previous models that incorporated more biological details \cite{JahTim15,MisKim16,CheSpre17,HagFuk18,NicClo19,MalBaz19,EckBag22} are limited in their capacity to provide clear and concise mechanistic explanations.
By contrast, our framework allowed us to systematically analyze the neuronal, synaptic and network mechanisms at work. We first corroborated the findings of \cite{RomTso15,TheRov18} as we recovered, in the limit of infinitely many neurons per population, an analogous deterministic regime of spontaneously emerging hippocampal replay patterns as in the Romani-Tsodyks (RT) model \cite{RomTso15}, see Fig.~\ref{fig:supp2}. Thanks to the similarity between our macroscopic model and the RT model, we are confident that our mesoscopic description readily allows for capturing realistic hippocampal dynamics not only on periodic tracks---which we considered here for simplicity and as a proof of concept---but also on linear tracks, in T-maze environments, and planar (and higher dimensional) fields; we leave these extensions for future work as well as the formation of theta sequences and phase precession \cite{TsoSka96,RomTso15,TheRov18}, see also \cite{BuzMos13,Col16}.
Our multiscale modeling framework directly links observed mesoscopic dynamics with the microscopic dynamics at the single neuron and single synapse level. This link could be a useful step towards a mechanistic understanding of the neuronal, synaptic and network constituents for generating spontaneous hippocampal activity critical for memory consolidation, recall and spatial working memory, navigational planning, as well as reward-based learning \cite{CarJad11,Fos17,OlaBus18,Pfe20}.

Importantly, our mesoscopic description opens a new perspective on the variability of hippocampal dynamics. In fact, we uncovered a novel, finite-sized induced metastable regime of hippocampal replay. At first glance, these fluctuation-induced quasi-traveling waves are similar in nature to those in \cite{RomTso15}. On time-scales of a few hundred milliseconds, both the deterministic RT model and our mesoscopic model exhibit spontaneously emerging bursts of activity that resemble a nonlocal replay event. As a single replay event unfolds, we showed that both models feature discontinuous replay trajectories---consistent with experimental observations \cite{PfeFos15,DenGil21,KraDru22}, which also underlines the biological plausibility of the ring-attractor assumption. On longer time-scales, however, our mesoscopic model allows for a much richer repertoire of replay dynamics. We found that the mesoscopic ring-attractor network can exhibit significantly larger variability for finite-sized induced replay dynamics than for deterministic, fatigue-induced dynamical regimes. This variability manifests in the spatio-temporally irregular succession of replay events (Figs.~\ref{fig:1Env}, \ref{fig:1EnvB} and \ref{fig:supp2}) and could be tested experimentally, following, e.g., \cite{AxmElg08,SchKal14,Buz15}. 
Consequently, our results insinuate that different replay dynamics can have distinct dynamical origins. In particular, replay dynamics in rodents are reportedly different during awake rest versus sleep \cite{McNSta21,KraDru22}, possibly relating to fluctuation- versus fatigue-induced metastability. 






\subsubsection*{Replay in brains and machines}
Our findings may also be relevant for human neuroscience, where a paradigm shift to decode cognition rather from off-task, than from task-based, neural activity seems to be imminent \cite{LiuNou22}. Hippocampal replay in rodents is a prime example for a ``representation-rich'' approach to spontaneous neural activity by uncovering the temporal structure of task-related representations. In understanding how the temporal dynamics of a particular neural activity pattern (e.g., a nonlocal traveling wave) unfolds, researchers could shed light on various cognitive functions that are subserved by spontaneous neural activity, including memory, learning, and decision-making \cite{JadKem12,OlaBus18}.
Recent technical advances in human neuroimaging have inspired ``human replay'' studies that investigate spontaneous task-related neural reactivations \cite{SchNiv19,LiuDol19,HigLiu21}, which bears strong resemblance to rodent replay. Instead of the spontaneous recall of an environment map in rodent hippocampal replay, the focus lies now on reactivation of a more abstract ``cognitive map'' of task space. 
As the associated cognitive processes include memory retrieval, planning and inference, and thus lie at the heart of sophisticated model-based reasoning, our results can also be regarded as a proof-of-concept for the model-based representation-rich approach advocated in \cite{LiuNou22} when re-interpreting the environmental maps stored in synaptic connectivities as cognitive maps.
Intriguingly, also in human replay studies, replay can occur in forward and backward direction, with putative functional roles for spatial and non-spatial learning \cite{KurEco16,LiuMat21}.
In particular, Liu and co-workers suggest that nonlocal backward replay may serve as a neural mechanism for model-based reinforcement learning \cite{LiuMat21}. A comprehensive view about the different roles the wide variety of replay dynamics may subserve, however, remains elusive.

Insights from machine learning, where replay is commonly implemented in artificial agents, may help to find answers about the putative computational functions \cite{WitChi21,HayKri21}.
``Experience replay'' was already introduced as a reinforcement learning technique in the early 90s \cite{Lin91} and is nowadays a crucial ingredient in building human-level intelligence in deep neural networks \cite{MniKav15,KumHas16}.
Note also that in reinforcement learning, a `model' has a similar meaning to the notion of a `cognitive map', which thus naturally bridges the gap from (human) cognition to artificial intelligence \cite{BehMul18}. 
Nonetheless, research on replay in neuroscience and machine learning has progressed largely in parallel, so that insights from the latter can also inform future neuroscientific studies.
An outstanding problem in the field of deep learning is the catastrophic forgetting problem in online learning \cite{MccCoh89,AbrRob05,HayKri21}. This problem is due to the fact that during online learning, data is not guaranteed to be independent and identically distributed ($i.i.d.$), which is a challenge for standard optimization methods. Replay-like methods are used to overcome this problem, but current replay implementations are computationally expensive to deploy. By contrast, uniform sampling of past experiences has proven to be remarkably efficient both in supervised learning \cite{ChaDok18,WuChe19, HayKaf20} and in reinforcement learning tasks \cite{MniKav13,MniKav15}. These insights provide some indirect, yet important evidence for the computational benefit of the stochastic replay that we have uncovered in this work and hint at an important role of the novel fluctuation-induced replay regime.



\subsubsection*{Biological limitations and extensions}

In this paper, we aimed at a minimal bottom-up population model that accounts for spiking noise, short-term synaptic plasticity and basic neuronal properties. The result can be regarded as a proof of concept that a simple nonlinear mesoscopic model, which enables the analysis of metastable dynamics in terms of network size and the aforementioned properties, can be derived from an underlying microscopic model. However, our model has several  biological limitations and lacks some important features. First, neurons exhibit post-spike \emph{refractoriness}, which is not captured by the LNP model. While the response of the instantaneous firing rate can be well reproduced by choosing the linear filter function $\kappa(t)$ and the nonlinearity $f(h)$ of the LNP model corresponding to realistic dynamical transfer functions of neurons with refractoriness \cite{OstBru11}, the temporal spike-train correlations caused by refractoriness violate the Poisson assumption of our derivation. Although the strict Poisson assumption can be relaxed to some degree \cite{SchGer20}, strong spike-train auto-correlations influence the noise properties of the mesoscopic model and hence the fluctuations of the population activities. This effect has already been described for the mesoscopic model of \cite{SchGer20} in the case of leaky integrate-and-fire neurons with pronounced refractoriness. Because our theory is based on the previous mesoscopic model, we expect that these effects carry over to the present model. How to account for non-Poisson statistics due to refractoriness in a mesoscopic theory with STP  is a challenging theoretical problem that is left for future research. We also mention that a related neuronal property is spike-triggered \emph{adaptation}, which -- similar to synaptic depression -- is a slow negative feedback mechanism. Incorporating adaptation into a mesoscopic theory, either instead of or in addition to depression, is interesting for two reasons: first, it represents an alternative slow fatigue mechanisms driving metastable dynamics, and second, adaptation is an important biological feature found in many cell types \cite{PozNau13}. A promising generalization of the present theory to adaptation could be based on the quasi-renewal approximation \cite{NauGer12} and its extension to mesoscopic theories \cite{DegSch14,SchDeg17}.

Apart from a more realistic description of neuronal properties, also the synaptic dynamics can be improved towards important biological features. First, the synaptic conductances exhibit temporal filtering and, in a conductance-based model, they enter the voltage dynamics in a multiplicative way. Both effects are neglected in our model. At least, the synaptic filtering of the conductance dynamics is straightforward to include in our theory as shown in \cite{SchGer20}. For an extension of the mesoscopic model to a conductance-based description of synaptic input, the interested reader is referred to the discussion of \cite{SchDeg17}. Second, the Tsodyks-Markram model considered here is a phenomenological and deterministic model of the synaptic dynamics. However, synaptic transmission is stochastic, and thus a stochastic STP model \cite{RosRub12} would be biologically more realistic (see also discussion in \cite{SchGer20}), but how to treat such stochasticity within a mesoscopic theory remains unclear.

A biologically difficult problem is how to account for realistic network topology.  Our theory applies to networks of multiple interacting homogeneous populations. In turn, each population is a fully-connected network of neurons. However, the connectivity of biological neural networks is not fully connected and often exhibits a large degree of heterogeneity. Regarding the first issue, we have shown previously that the full connectivity model represents an effective model that faithfully reproduces the dynamics of a non-fully-connected, random network with fixed in-degree if the synaptic weights are rescaled correspondingly \cite{SchGer20} (see also \cite{SchDeg17} for the case of static synapses). The second issue is a principle challenge for mean-field theories as they rely on the possibility to use averages over many neurons, and hence cannot describe heterogeneous networks that are strongly affected by single neurons. However, in many cases it might be valid to subdivide the network into many small subpopulations that can be regarded as roughly homogeneous. Following this strategy, it is crucial to have a mesoscopic description of the subpopulations because the grouping of neurons with similar properties may result in small population sizes. For example, in our hippocampal network model, we have grouped place cells with highly overlapping place fields into one homogeneous subpopulation. If the number of these similarly tuned place cells is small, the mesoscopic framework can show its full strength because in this case, the jump-diffusion model still provides an accurate description (see Fig.~\ref{fig:updown} for $N=30$ neurons). Finally, we note that a basic type of network heterogeneity in biology, the separation of excitatory and inhibitory neurons (Dale's law), is not realized  in our hippocampal network with ``Mexican-hat''-type connectivity. However, it has been shown that such connectivity can be re-implemented in accordance with Dale's law by two layers of neurons, one excitatory and one inhibitory layer \cite{OzeFin09}.

\subsubsection*{Theoretical challenges}

The diffusion model Eq.~\eqref{eq:meso} and its derivation based on temporal coarse-graining \cite{Gil00} greatly simplify our previous theoretical work \cite{SchGer20} but both results are consistent as shown below in Methods. The simplicity of our new derivation and the remarkable agreement of the diffusion model with microscopic simulations beg the question of whether the diffusion model is the \textit{exact} diffusion approximation \cite{Kur78}. The only way to give a positive answer to this question would be to propose a rigorous diffusion approximation proof (as in \cite{DitLoe17} for the case of LNP neurons without STP), which we leave as an open mathematical problem. A second open theoretical question is the convergence of the multi-population model Eq.~\eqref{eq:ring_meso} to a stochastic neural field equation. The circular environment we study can be regarded as a space-discretized stochastic neural field but the exact expression of the continuous equation and its physical interpretation is unclear and will be subject of future work. Ideally, one would want to be able to prove the convergence to such an equation, as in \cite{CheOst20} in the case without STP. 

The low-dimensional mesoscopic model could be also interesting from a data-analytical perspective. Recently, Bayesian state-space models have been developed to infer replay events from spiking data \cite{DenGil21}. Such inference does not exploit any knowledge about the dynamical mechanisms underlying the data and the likelihood function is assumed ad hoc. Our low-dimensional mesoscopic could provide an analytical likelihood function so as to enable improved data assimilation methods to infer replay events. 

\paragraph*{}

In conclusion, we have put forward a multiscale framework for systematically investigating metastable network dynamics in finite-sized networks of LNP-STP neurons using a bottom-up mesoscopic model. This model is efficient to analyze and simulate and is also versatile for incorporating more biological realism. Thanks to a unique link between the underlying microscopic network and its mesoscopic description, it becomes possible to disentangle the differential roles of neuronal, synaptic and network properties---in particular the network size---for emerging metastable brain dynamics. The mesoscopic model may also be instrumental for distinguishing between fatigue-driven and fluctuation-driven metastability because of their distinct statistical predictions---as in the case of hippocampal replay. Such predictions could be tested experimentally and reveal the dynamical origin of spontaneous neural activity.

\section*{Methods}

\subsection*{Diffusion approximation for the mesoscopic dynamics with short-term depression}

The microscopic dynamics of a network of $N$ LNP spiking neurons with short-term synaptic depression and exponential linear filter dynamics are given by Eq.~\eqref{eq:micro}. To derive the diffusion model of the mesoscopic dynamics Eq.~\eqref{eq:meso}, we focus on the mesoscopic variables $h(t), x(t)$ and $Q(t)$ defined as  in Eq.~\eqref{eq:empirical}:
\begin{equation*}
    h(t):=\frac{1}{N}\sum_{i=1}^N h_i(t), \quad x(t):=\frac{1}{N}\sum_{i=1}^N x_i(t) \quad \text{and} \quad Q(t) := \frac{1}{N}\sum_{i=1}^N x_i^2(t).
\end{equation*}
From Eq.~\eqref{eq:micro} and using the definition of $x(t)$, we get
\begin{subequations}
\begin{align}
    \od{h}{t}&=\frac{\mu(t)-h}{\tau}+JU_0\frac{1}{N}\sum_{i=1}^Nx_i(t^-)s_i(t), \\
    \od{x}{t}&=\frac{1-x}{\tau_D}-U_0\frac{1}{N}\sum_{i=1}^N x_i(t^-)s_i(t).
\end{align}
\end{subequations}
To approximate the sum $\frac{1}{N}\sum_{i=1}^Nx_i(t^-)s_i(t)$ by a diffusion term which only involves mesoscopic variables, we follow the coarse-graining approach by Gillespie \cite{Gil00} for the derivation of a ``chemical Langevin equation''.
To this end, we study the stochastic increments $\int_t^{t+\Delta t} \frac{1}{N}\sum_{i=1}^Nx_i(\hat{t}^-)s_i(\hat{t})d\hat{t}$, where $\Delta t > 0$ is assumed to be a \textit{macroscopically infinitesimal} time step \cite{Gil00}: $\Delta t$ is small enough such that (i) the $x_i$'s can be assumed to jump at most once in the time interval $[t,t+\Delta t]$, (ii) $x_i(\tau_i^-) \approx x_i(t^-)$ if neuron $i$ has a spike at time $\tau_i\in[t,t+\Delta t]$, and (iii) $h(\hat{t})\approx h(t^-)$ for all $\hat{t}\in[t,t+\Delta t]$; and  $\Delta t$ is large enough such that many neurons spike in the time interval $]t,t+\Delta t]$. These assumptions are expected to hold if $\Delta t\ll \tau,\tau_D$ and $1\ll Nf(h(t))\Delta t\ll N$ for all $t$. By the smallness assumption, we have
\begin{equation}\label{eq:increments}
    \int_t^{t+\Delta t} \frac{1}{N}\sum_{i=1}^Nx_i(\hat{t}^-)s_i(\hat{t})d\hat{t} \approx \frac{1}{N}\sum_{i=1}^N x_i(t^-)z_i(t),
\end{equation}
where $\{z_i(t)\}_{i=1}^N$ are \textit{i.i.d.} Bernoulli random variables with mean $f(h(t^-))\Delta t$. Conditioned on $h(t^-)$, all the variables $x_1(t^-), \dots, x_N(t^-)$,  $z_1(t), \dots, z_N(t)$ are independent, and the $\{x_i(t^-)\}_{i=1}^N$ are \textit{i.i.d.}. Hence, by the Central Limit Theorem,
\begin{equation*}
    \frac{\sum_{i=1}^N x_i(t^-)z_i(t) - N\mathbb{E}[x_1(t^-)z_1(t)\,|\,h(t^-)]}{\sqrt{N}} \xrightarrow[N\to\infty]{\mathcal{L}} \mathcal{N}\left(0,\text{Var}[x_1(t^-)z_1(t)\,|\,h(t^-)]\right).
\end{equation*}
We now use the empirical averages Eq.~\eqref{eq:empirical} to approximate the conditional expectation and variance:
\begin{align*}
\mathbb{E}[x_1(t^-)z_1(t)\,|\,h(t^-)] &= \mathbb{E}[x_1(t^-)\,|\,h(t^-)]f(h(t^-))\Delta t \approx x(t^-) f(h(t^-))\Delta t\\
    \text{Var}[x_1(t^-)z_1(t)\,|\,h(t^-)] &=\expect{x_1^2(t^-)\middle|h(t^-)}\expect{z_1^2(t^-)\middle|h(t^-)}\\
    &\qquad -\expect{x_1(t^-)\middle|h(t^-)}^2\expect{z_1(t^-)\middle|h(t^-)}^2\\
    &\approx Q(t^-)f(h(t^-))\Delta t + O(\Delta t^2).
\end{align*}
We can now approximate the increment Eq.~\eqref{eq:increments} by a Gaussian:
\begin{equation*}
    \int_t^{t+\Delta t} \frac{1}{N}\sum_{i=1}^Nx_i(\hat{t}^-)s_i(\hat{t})d\hat{t} \,\sim\, \mathcal{N}\left(x(t^-)f(h(t^-))\Delta t,\, \frac{Q(t^-)f(h(t^-))\Delta t}{N}\right).
\end{equation*}
Taking the limit $\Delta t \to 0$, we obtain the diffusion approximation
\begin{subequations}\label{eq:diffusion}
\begin{align}
    \od{h}{t}&=\frac{\mu(t)-h}{\tau}+JU_0xf(h) + JU_0 \sqrt{\frac{Q(t)f(h)}{N}}\xi(t), \\
    \od{x}{t}&=\frac{1-x}{\tau_D}-U_0xf(h) - U_0\sqrt{\frac{Q(t)f(h)}{N}}\xi(t),
\end{align}
\end{subequations}
where $\xi$ is a Gaussian white noise with auto-correlation function $\langle\xi(t)\xi(t')\rangle=\delta(t-t')$. 

To close the system Eq.~\eqref{eq:diffusion}, we have to derive the dynamics of $Q(t)$.
Going back to Eq.~\eqref{eq:micro_x}, by Itô's formula for jump processes, we have
\begin{equation*}
    \od{x_i^2}{t} = 2\frac{x_i - x_i^2}{\tau_D} - U_0(2-U_0)x_i^2(t^-)s_i(t).
\end{equation*}
Taking the empirical average, we get
\begin{equation} \label{eq:Q_emp}
    \od{Q}{t} = 2\frac{x - Q}{\tau_D} - U_0(2-U_0)\frac{1}{N}\sum_{i=1}^N x_i(t^-)^2s_i(t).
\end{equation}
We could follow the same steps as before and try to obtain a diffusion approximation for Eq.~\eqref{eq:Q_emp}. However, the fluctuations of such a diffusion approximation would be of order $1/\sqrt{N}$ and since $Q(t)$ affects the dynamics of Eq.~\eqref{eq:diffusion} only through the term $\sqrt{Q(t)f(h)/N}\xi(t)$, the effect of the fluctuation of $Q(t)$ on $h(t)$ and $x(t)$ are of order $N^{-3/2}$ and can therefore be neglected when $N$ is large. Hence, we approximate the increments by their (approximate) expectation: for the time step $\Delta t > 0$,
\begin{equation*}
    \int_t^{t+\Delta t}\frac{1}{N}\sum_{i=1}^N x_i(\hat{t}^-)^2s_i(\hat{t})d\hat{t} \,\approx\, Q(t^-)f(h(t^-))\Delta t,
\end{equation*}
whence,
\begin{equation} \label{eq:Q}
    \od{Q}{t} = 2\frac{x - Q}{\tau_D} - U_0(2-U_0)Qf(h).
\end{equation}
Finally, gathering Eqs.~\eqref{eq:diffusion} and \eqref{eq:Q}, we obtain the mesoscopic dynamics in form of the diffusion model Eq.~\eqref{eq:meso}.

The fact that we are here considering the diffusion approximation (or Langevin dynamics) allows to significantly shorten the original derivation of the mesoscopic model presented in \cite{SchGer20}. In particular, in the present derivation, we do not need to approximate the distribution of the $x_i(t^-)$'s in Eq.~\eqref{eq:increments} by a Gaussian, since we only need to approximate the sum in Eq.~\eqref{eq:increments} by a Gaussian. Note that the arguments enabling the present derivation were already hinted at in the Section ``Remarks on the approximation'' and Appendix B of \cite{SchGer20} but not put together. Both derivations lead to the same mesoscopic model, except that here, for simplicity, we neglect fluctuations of order $N^{-3/2}$ and we consider the diffusion limit; see also the subsection ``Reduction to a pure diffusion process'' below for an explicit derivation of the diffusion model from the original mesoscopic model presented in \cite{SchGer20}.

\subsection*{Jump-diffusion model  with synaptic depression and facilitation}

Starting from  our previous work \cite{SchGer20}, we can derive an improved mesoscopic model that also accounts for synaptic facilitation and the shot-noise character of the finite-size spiking noise. 
When allowing only for short-term depression while keeping the facilitation variable constant, the resulting mesoscopic dynamics boils down to the jump-diffusion model Eq.~\eqref{eq:model-depress-3} with hybrid noise. For large $N\gg 1$, the Poisson shot noise can be simplified under a diffusion approximation, which recovers the diffusion model Eq.~\eqref{eq:meso} derived in the previous section.

\subsubsection*{Microscopic model with synaptic depression and facilitation}
\label{sec:micro-full}

We consider a network of LNP neurons with dynamic synapses similar to Eq.~\eqref{eq:micro} but now complemented with a facilitation variable $\hat{u}_i$ for each neuron $i = 1,\dots,N$. The full synaptic dynamics corresponds to the STP model by Tsodyks and Markram \cite{TsoPaw98,MonBar08} and results in the following microscopic network model:
\begin{subequations}
\label{eq:micro-full-stp}
\begin{align}
  r_i(t)&=f\lrrund{\hat{h}_i(t^-)}\\
  \frac{d\hat{h}_i}{dt}&=\frac{\mu(t)-\hat{h}_i}{\tau}+\frac{J}{N}\sum_{j=1}^N\hat{u}_j(t^-)\hat{x}_j(t^-)s_j(t) \label{eq:micro-h}\\
  \frac{d\hat{u}_j}{dt}&=\frac{U_0-\hat{u}_j}{\tau_F}+U(1-\hat{u}_j(t^-))s_j(t)\\
  \frac{d\hat{x}_j}{dt}&=\frac{1-\hat{x}_j}{\tau_D}-\hat{u}_j(t^-)\hat{x}_j(t^-)s_j(t),
\end{align}
\end{subequations}
where $s_i(t)=\sum_k\delta(t-t_k^i)$ is a point process with conditional intensity $r_i(t)$.
In Eq.~\eqref{eq:micro-full-stp}, $\tau_F$ and $\tau_D$ are the facilitation and depression time constants, respectively, and $U_0$ is the
baseline utilization of synaptic resources, whereas $U$ determines the increase in the utilization of
synaptic resources by a spike. 
As before, $\hat{u}_j(t^-)$ is a shorthand for the left limit at time $t$.
We note that, for simplicity, we only consider the case of full connectivity. However, as shown in our previous work \cite{SchGer20}, the mean-field theory also works well for random connectivity with fixed in-degree.

We remark that the model Eq.~\eqref{eq:micro-h} corresponds to LNP neurons with an exponential linear filter, which we have chosen in the Results part for simplicity. The mesoscopic theory developed here can readily be extended to LNP neurons described by a general linear filter $\kappa(t)$. In this case and assuming $\hat{h}_i(0)=0$, the $h$-dynamics will be given by 
\begin{equation}
\hat{h}_i(t)=\int_0^{t^+} \kappa(t-t')\lrrund{\frac{\mu(t')}{\tau}+\frac{J}{N}\sum_{j=1}^N \hat{u}_j(t'^-)\hat{x}_j(t'^-)s_j(t')}\,dt'.
\end{equation}
 With this extension more realistic neuronal dynamics can be modeled \cite{OstBru11}. The simple dynamics Eq.~\eqref{eq:micro-h} is recovered if the linear filter is chosen as $\kappa(t)=e^{-t/\tau}\theta(t)$, where $\theta(t)$ is the Heaviside step function.

\subsubsection*{Mesoscopic model}
\label{sec:meso-full}

In the Appendix B of \cite{SchGer20}, it has been shown that the mesoscopic dynamics of the empirical variables 
\begin{equation} \label{eq:empirical2}
\begin{gathered}
h(t):= \frac{1}{N}\sum_{i=1}^N \hat{h}_i(t), \quad u(t):= \frac{1}{N}\sum_{i=1}^N \hat{u}_i(t), \quad x(t) := \frac{1}{N}\sum_{i=1}^N \hat{x}_i(t),\\
P(t):= \frac{1}{N}\sum_{i=1}^N \hat{u}_i^2(t), \quad Q(t):= \frac{1}{N}\sum_{i=1}^N \hat{x}_i^2(t), \quad R(t):= \frac{1}{N}\sum_{i=1}^N \hat{u}_i(t) \hat{x}_i(t)
\end{gathered}
\end{equation}
can be approximated in discrete time with a macroscopic infinitesimal time step $\Delta t$ (as defined above) by the moment-closure equations
  \begin{subequations}
  \label{eq:update}
  \begin{align}
  h_{k+1}&=h_k+\frac{\mu_k-h_k}{\tau}\Delta t+\frac{J}{N}\left[R_k\Delta n_k +(u_k\varepsilon_k^x + x\varepsilon_k^u)\sqrt{\Delta n_k}\right], \\
    u_{k+1}&=u_k+\frac{U_0-u_k}{\tau_F}\Delta t+\frac{U}{N}\left[(1-u_k)\Delta n_k - \varepsilon_k^u\sqrt{\Delta n_k}\right]\\
    x_{k+1}&=x_k+\frac{1-x_k}{\tau_D}\Delta t-\frac{1}{N}\left[R_k\Delta n_k +(u_k\varepsilon_k^x + x\varepsilon_k^u)\sqrt{\Delta n_k}\right], \\
    P_{k+1}&=P_k+2 \frac{U_0u_k-P_k}{\tau_F}\Delta t+\frac{1}{N}\lreckig{\mu^P(u_k)\Delta n_k +\varepsilon_k^P\sqrt{\Delta n_k}},\\
    Q_{k+1}&=Q_k+2 \frac{x_k-Q_k}{\tau_D}\Delta t+\frac{1}{N}\lreckig{\mu^Q(u_k,x_k,P_k,Q_k,R_k)\Delta n_k +\varepsilon_k^Q\sqrt{\Delta n_k}},\\
    R_{k+1}&=R_k+\frac{U_0x_k-R_k}{\tau_F}\Delta t+\frac{u_k-R_k}{\tau_D}\Delta t+\frac{1}{N}\lreckig{\mu^R(u_k,x_k,P_k,R_k)\Delta n_k +\varepsilon_k^R\sqrt{\Delta n_k}},
  \end{align}
  \end{subequations}
where $u_k = u(k \Delta t)$ and analogous expressions hold for $h_k,x_k, P_k, Q_k,R_k$ and the external input $\mu_k$. In Eq.~\eqref{eq:update}, we use the abbreviations
\begin{align*}
  \mu^P(u) &= U \big(P (U - 2) - 2 u (U - 1) + U\big),\\
  \mu^Q(u,x,P,Q,R) &= P Q - 2 Q u + 2 \big(R + (u - 2) x\big) (R - u x),\\
  \mu^R(u,x,P,R) &= (U (1 - u)^2 - u^2) x + (U - 1) x (P - u^2) + 2 (U (u - 1) - u) (R - u x),
\end{align*}
and
\begin{align*}
  \varepsilon_k^P &= 2U \big(1 + u_k (U - 2) - U\big)\varepsilon_k^u, \\
  \varepsilon_k^Q &= 2(u_k - 1)x_k^2\varepsilon_k^u + 2u_k(u_k-2)x_k\varepsilon_k^x,\\
  \varepsilon_k^R &= 2\big(U(u_k - 1) - u_k\big)x_k\varepsilon_k^u + \big(U(1 - u_k)^2 - u_k^2\big)\varepsilon_k^x.
\end{align*}
Importantly, the dynamics is driven by two sources of noise: First, $\varepsilon_k^u$ and $\varepsilon_k^x$ are correlated Gaussian random numbers with means $\langle\varepsilon_k^u\rangle=\langle\varepsilon_k^x\rangle=0$ and (co)variances
\begin{equation}
    \langle\varepsilon_k^u\varepsilon_l^u\rangle=(P_k-u_k^2)\delta_{k,l},\quad \langle\varepsilon_k^x\varepsilon_l^x\rangle=(Q_k-x_k^2)\delta_{k,l},\quad \langle\varepsilon_k^u\varepsilon_l^x\rangle=(R_k-u_kx_k)\delta_{k,l},
  \end{equation}
  where $\delta_{k,l}$ is the Kronecker delta.
The random numbers $\varepsilon_k^u$ and $\varepsilon_k^x$ reflect the heterogeneity of $\hat{u}_i$ and $\hat{x}_i$ across synapses $i=1,\dotsc,N$, respectively. Second, $\Delta n_k$ represents the total spike count in the time step $\Delta t$ which is drawn independently from a Poisson distribution with mean $Nf(h_k)\Delta t$:
\begin{equation}
    \Delta n_k\sim\text{Pois}[Nf(h_k)\Delta t].
\end{equation}
This equation closes the discrete mesoscopic dynamics with STP derived in \cite{SchGer20}. 

We will now use the discrete dynamics to derive a Langevin equation in continuous time.
Because of our assumption that $\Delta t$ is large enough such that it contains many spikes, i.e. $\langle\Delta n_k\rangle\equiv Nf(h_k)\Delta t\gg 1$, we can use again a Gaussian approximation. Thus, we write $\Delta n_k\approx Nf(h_k)\Delta t+\sqrt{Nf(h_k)}\Delta W_k$, where $\Delta W_k$ is an independent, normally distributed random number with $\langle\Delta W_k\rangle=0$ and $\langle{\Delta W_k^2}\rangle=\Delta t$.  Furthermore, the noise terms appearing in Eq.~\eqref{eq:update} can be written within a Gaussian approximation as
\begin{equation}
  \varepsilon_k^u\sqrt{\Delta n_k}\approx\sqrt{N(P-u^2)f(h_k)}\Delta W_k^u,\qquad \varepsilon_k^x\sqrt{\Delta n_k}\approx\sqrt{N(Q-x^2)f(h_k)}\Delta W_k^x,
\end{equation}
where we neglected terms of order $\mathcal{O}(N^{\frac{1}{4}})$ and where $\Delta W_k^u$ and $\Delta W_k^x$ are mean-zero Gaussian random numbers with covariance
\begin{equation}
  \langle\Delta W_k^u\Delta W_l^u\rangle=\langle\Delta W_k^x\Delta W_l^x\rangle=\delta_{k,l}\Delta t,\qquad \langle \Delta W_k^u\Delta W_l^x\rangle=\delta_{k,l}\rho_k\Delta t
\end{equation}
Here, we introduced the correlation coefficient
\begin{equation}
  \rho_k=\frac{R_k-u_kx_k}{\sqrt{(P_k-u_k^2)(Q_k-x_k^2)}}.
\end{equation}
It follows that, within the Gaussian approximation, the other noise terms are given by
\begin{align*}
  \varepsilon_k^P\sqrt{\Delta n_k} &\approx 2U \big(1 + u_k (U - 2) - U\big)\sqrt{N(P-u^2)f(h_k)}\Delta W_k^u, \\
  \varepsilon_k^Q\sqrt{\Delta n_k} &\approx 2(u_k - 1)x_k^2\sqrt{N(P-u^2)f(h_k)}\Delta W_k^u\\
  &\quad+ 2u_k(u_k-2)x_k\sqrt{N(Q-x^2)f(h_k)}\Delta W_k^x,\\
  \varepsilon_k^R\sqrt{\Delta n_k} &\approx 2\big(U(u_k - 1) - u_k\big)x_k\sqrt{N(P-u^2)f(h_k)}\Delta W_k^u\\
  &\quad + \big(U(1 - u_k)^2 - u_k^2\big)\sqrt{N(Q-x^2)f(h_k)}\Delta W_k^x.
\end{align*}
Taking the continuum limit $\Delta t\rightarrow 0$ yields the Itô stochastic differential equation
\begin{subequations}
  \begin{align}
  dh_t&=\frac{\mu_t-h_t}{\tau}dt+\frac{J}{N}\left[R_tdn_t +u_t\sqrt{N(Q_t-x_t^2)f(h_t)}dW_t^x + x_t\sqrt{N(P_t-u_t^2)f(h_t)}dW_t^u\right], \\
    du_t&=\frac{U_0-u_t}{\tau_F}dt+\frac{U}{N}\left[(1-u_t)dn_t - \sqrt{N(P_t-u_t^2)f(h_t)}dW_t^u\right]\\
    dx_t&=\frac{1-x_t}{\tau_D}dt-\frac{1}{N}\left[R_tdn_t +u_t\sqrt{N(Q_t-x_t^2)f(h_t)}dW_t^x + x_t\sqrt{N(P_t-u_t^2)f(h_t)}dW_t^u\right], \\
    dP_t&=2 \frac{U_0u_t-P_t}{\tau_F}dt+\frac{1}{N}\lreckig{\mu^P(u_t)dn_t +2U \big(1 + u_t (U - 2) - U\big)\sqrt{N(P_t-u_t^2)f(h_t)}dW_t^u},\\
    dQ_t&=2 \frac{x_t-Q_t}{\tau_D}dt+\frac{1}{N}\bigg[\mu^Q(u_t,x_t,P_t,Q_t,R_t)dn_t \nonumber\\
    &\quad+2(u_t - 1)x_t^2\sqrt{N(P_t-u_t^2)f(h_t)}dW_t^u+ 2u_t(u_t-2)x_t\sqrt{N(Q_t-x_t^2)f(h_t)}dW_t^x\bigg],\\
    dR_t&=\frac{U_0x_t-R_t}{\tau_F}dt+\frac{u_t-R_t}{\tau_D}dt+\frac{1}{N}\bigg[\mu^R(u_t,x_t,P_t,R_t)dn_t\nonumber\\
        &\quad+2\big(U(u_t - 1) - u_t\big)x_t\sqrt{N(P_t-u_t^2)f(h_t)}dW_t^u\nonumber\\
    &\quad+ \big(U(1 - u_t)^2 - u_t^2\big)\sqrt{N(Q_t-x_t^2)f(h_t)}dW_t^x\bigg].
  \end{align}
  with Poisson noise
  \begin{equation}
    dn_t=\pi(dt,[0,Nf(h_{t^-})]),
  \end{equation}
 \end{subequations}
  where $\pi$ is a two-dimensional Poisson random measure with mean $\langle\pi(ds,dt)\rangle=dsdt$ (i.e. $n_t$ is a counting process with stochastic intensity $Nf(h_{t^-})$ and $dn_t/dt$ is the associated Dirac delta spike train).
  Furthermore, $W_t^u$ and $W_t^x$ are Wiener processes, where $W_t^u$ and $W_t^x$ have correlated increments
\begin{subequations}
\begin{gather}
  \langle dW_t^udW_s^u\rangle=\langle dW_t^xdW_s^x\rangle=\delta(t-s)dtds,\\
  \langle dW_t^udW_s^x\rangle=\frac{R_t-u_tx_t}{\sqrt{(P_t-u_t^2)(Q_t-x_t^2)}}\delta(t-s)dtds.
\end{gather}
\end{subequations}
Introducing the Gaussian white-noise processes
\begin{subequations}
\begin{align}
\xi_x(t)&=\sqrt{\frac{(Q_t-x_t^2)f(h_t)}{N}}\frac{dW_t^x}{dt},\\ 
\xi_u(t)&=\sqrt{\frac{(P_t-u_t^2)f(h_t)}{N}}\frac{dW_t^u}{dt},
\end{align} 
\end{subequations}
the full stochastic dynamics of the mesoscopic neural-mass model can be rewritten in the form of a Langevin equation
\begin{subequations}
    \label{eq:langevin}
  \begin{align}
  \frac{dh}{dt}&=\frac{\mu(t)-h}{\tau}+J\lreckig{RA(t) +u\xi_x(t) + x\xi_u(t)}, \label{eq:h-meso-full-exp}\\
    \frac{du}{dt}&=\frac{U_0-u}{\tau_F}+U(1-u)A(t) - U\xi_u(t)\\
    \frac{dx}{dt}&=\frac{1-x}{\tau_D}-RA(t) -u\xi_x(t) - x\xi_u(t), \\
    \frac{dP}{dt}&=2 \frac{U_0u-P}{\tau_F}+\mu^P(u)A(t) +2U \big(1 + u(U - 2) - U\big)\xi_u(t),\\
    \frac{dQ}{dt}&=2 \frac{x-Q}{\tau_D}+\mu^Q(u,x,P,Q,R)A(t)  + 2(u-1)x^2\xi_u(t) + 2u(u-2)x\xi_x(t)\\
    \frac{dR}{dt}&=\frac{U_0x-R}{\tau_F}+\frac{u-R}{\tau_D}+\mu^R(u,x,P,R)A(t)
        +2\big(U(u - 1) - u\big)x\xi_u(t)+ \big(U(1 - u)^2 - u^2\big)\xi_x(t).
  \end{align}
  with
  \begin{equation}
    A(t)=\frac{1}{N}\frac{dn_t}{dt}.
  \end{equation}
  The Gaussian white noise processes are given by their covariance functions
  \begin{align}
    \langle \xi_u(t)\xi_u(s)\rangle&=\frac{(P_t-u_t^2)f(h_t)}{N}\delta(t-s),\\
  \langle \xi_x(t)\xi_x(s)\rangle&=\frac{(Q_t-x_t^2)f(h_t)}{N}\delta(t-s),\label{eq:langevin_xix}\\
  \langle \xi_u(t)\xi_x(s)\rangle&=(R_t-u_tx_t)\delta(t-s).
\end{align}
\end{subequations}
Equation \eqref{eq:langevin} constitutes the jump-diffusion model for the full Tsodyks-Markram STP model with depression and facilitation.

In the case of a general impulse-response function $\kappa$, the derivation of the mesoscopic STP dynamics does not change. The only difference is that the dynamics for $h$ above needs to be changed to corresponding convolution equations. Therefore, for a general impulse-response function one only needs to replace Eq.~\eqref{eq:h-meso-full-exp} by
\begin{equation}
\label{eq:h-meso-full-kappa}
    h(t)=\int_0^{t^{+}}\kappa(t-\tl)\left\{\frac{\mu(\tl)}{\tau}+J\lreckig{R(\tl^{-})A(\tl)+u(\tl^{-})\xi_x(\tl)+x(\tl^{-})\xi_u(\tl)}\right\}\,d\tl.
\end{equation}

\subsection*{Mesoscopic model with pure synaptic depression}
In order to obtain the jump-diffusion model Eq.~\eqref{eq:model-depress-3} only with short-term synaptic depression but without facilitation as considered in the Results section, we set $u\equiv U_0, P \equiv U_0^2$ and $R \equiv U_0x$. Then, $\mu^Q(u,x,P,Q,R) = -U_0(2-U_0)Q$ and $\xi_u(t) \equiv 0$, and the dynamics Eq.~\eqref{eq:langevin} reduce to
\begin{subequations}
    \label{eq:langevin_depress}
  \begin{align}
    \frac{dx}{dt}&=\frac{1-x}{\tau_D}-U_0xA(t) -U_0\xi_x(t) , \\
    \frac{dQ}{dt}&=2 \frac{x-Q}{\tau_D}-U_0(2-U_0)Qf(h(t)) \label{eq:Q-full} ,
  \end{align}
  \end{subequations}
  with the Gaussian white noise process $\xi_x(t)$ defined as in Eq.~\eqref{eq:langevin_xix}.
In Eq.~\eqref{eq:Q-full}, we have neglected the noise terms and replaced the population activity $A(t)$ by its mean $f(h(t))$ because they enter the dynamics of $x$ only to order $N^{-3/2}$.
Finally, introducing the new mesoscopic variable $\tilde Q = Q-x^2$, we can combine Eqs.~\eqref{eq:langevin_depress} and \eqref{eq:langevin_xix} to arrive at the jump-diffusion model Eq.~\eqref{eq:model-depress-3}.

Again, in the case of a general impulse-response function $\kappa$, the stochastic differential equation \eqref{eq:jump-diff-depress-h} for $h(t)$ needs to be replaced by the corresponding integral expression
\begin{equation}
    h(t)=\int_0^{t^{+}}\kappa(t-\tl)\left\{\frac{\mu(\tl)}{\tau}+JU_0\lreckig{x(\hat{t}^-)A(\hat{t})+\sqrt{\frac{\tilde Q f(h)}{N}}\xi_x(\hat{t})}\right\}\,d\tl.
\end{equation}

\subsubsection*{Reduction to a pure diffusion process}
We can further reduce the jump-diffusion model Eq.~\eqref{eq:model-depress-3} in the large $N\gg 1$ limit by exploiting the Gaussian approximation of the Poisson shot noise Eq.~\eqref{eq:popact} representing the empirical population activity
\begin{equation}
A(t) = \frac1N \frac{n(t)}{dt} \approx f\big(h(t)\big)  + \xi_A(t),
\end{equation}
where the Gaussian white noise process $\xi_A(t)$ has the auto-correlation function 
\begin{equation}
\langle \xi_A(t)\xi_A(s)\rangle = \frac{f(h(t))}N \delta(t-s).
\end{equation}
Consequently, we find that
\begin{align*}
U_0xA(t) + U_0\xi_x(t) = U_0xf(h) + U_0 \big[ x\xi_A(t) + \xi_x(t)\big].
\end{align*}
We can simplify the term in brackets by capitalizing on the fact that $x\xi_A$ and $\xi_x$ are independent Gaussian white noises, whose sum is itself a Gaussian random variable with variance
\begin{align*}
\frac{x^2 f(h)}{N} + \frac{(Q-x^2) f(h)}{N} = \frac{Q f(h)}{N}.
\end{align*}
Finally, we can replace the term $U_0xA(t) + U_0\xi_x(t) $ in the $h$- and $x$-dynamics of the jump-diffusion model by
\begin{equation}
U_0xA(t) + U_0\xi_x(t) = U_0x f\big( h(t)\big) + U_0\frac{Q f\big( h(t)\big)}{N} \xi(t),
\end{equation}
where $\xi(t)$ is a Gaussian white noise with auto-correlation function $\langle \xi(t)\xi(s) \rangle = \delta(t-s)$, and we retrieve, in an alternative way, the mesoscopic diffusion model Eq.~\eqref{eq:meso} with short-term synaptic depression.

As before, for a general impulse-response function $\kappa$, the stochastic differential equation \eqref{eq:diff-model-h} for $h(t)$ needs to be replaced by the corresponding integral expression
\begin{equation}
    h(t)=\int_0^{t^{+}}\kappa(t-\tl)\left\{\frac{\mu(\tl)}{\tau}+JU_0\lreckig{x(\tl)f(h(\tl)) + \sqrt{\frac{Q(\tl^{-})f(h(\tl^{-}))}{N}}\xi(\tl)}\right\}\,d\tl.
\end{equation}

\subsection*{Recurrent network parameters and numerical simulations}

In the Results section, we presented and analyzed the network dynamics of a single excitatory population consisting of $N$ LNP-STD spiking neurons following the microscopic dynamics Eq.~\eqref{eq:micro} or the mesoscopic dynamics Eq.~\eqref{eq:meso}/\eqref{eq:model-depress-3}. For the ring-attractor network model to study hippocampal replay dynamics, we used the microscopic dynamics Eq.~\eqref{eq:ring_micro} and the mesoscopic dynamics Eq.~\eqref{eq:ring_meso}.
The parameters to obtain Figures~\ref{fig:meta}--\ref{fig:3env} are detailed in Table~\ref{table:parameters}.
The number $N$ of neurons per population $\alpha=1,\dots,M$ is indicated inside the Figures. 
The model specification and parameters of the Romani-Tsodyks (RT) model for fatigue-induced hippocampal replay, with which we compared our results for the mesoscopic ring-attractor network model in the single and in the multiple environment case in Figs.~\ref{fig:1Env}--\ref{fig:3env}, respectively, can be found in \cite{RomTso15}.

\begin{table}[t]
\begin{adjustwidth}{-0.in}{0in} 
\centering
\caption{
{\bf Parameters used in Figs.~\ref{fig:meta}--\ref{fig:3env}}}
\begin{tabular}{|l c c+c|c|c|c|}
\hline
{\bf Parameters} & &{\bf Fig.} & {\bf \ref{fig:meta}} & {\bf \ref{fig:updown}} & {\bf \ref{fig:1Env}/\ref{fig:1EnvB}} & {\bf\ref{fig:3env}} \\ \thickhline
Synaptic time constant & $\tau$ & [s] & \hspace{.1cm}0.05\hspace{.1cm} & \hspace{.1cm}0.05\hspace{.1cm} & \hspace{.1cm}0.01\hspace{.1cm} & \hspace{.1cm}0.01\hspace{.1cm} \\
Depression time constant & $\tau_D$ & [s] & 0.8 & 0.6 & 0.8 & 0.8\\
Utilization of synaptic resources & $U_0$ & & 0.4 & 0.4 & 0.8 & 0.8\\
Transfer function $f(h)$ & & & & & &\\
---Suprathreshold slope & $r$ & & 3.15 & 3.15 & 1.0 & 1.0\\
---Smoothness at threshold & $\alpha$&  & 0.25 & 0.2 & 1.0 & 1.0\\
---Threshold & $h_0$ & [mV] & 2.0 & 2.0 & 0.0 & 0.0\\
Coupling constant & $J \cdot \tau$ & [mV] & 3.5 & 3.5 & -- & --\\
Uniform feedback inhibition & $J_0 \cdot \tau$ & [mV] & -- & -- & 13 & 16\\
Map-specific interaction & $J_1 \cdot \tau$ & [mV] & -- & -- & 30 & 25\\
External input (meso-,microscopic models) & $\mu$ & [mV] & 1.4 & 1.4 & -1.4 & -1.5\\
External input (only macroscopic model) & $\mu_\text{macro}$ & [mV] & -- & -- & -0.9 & -0.35\\
Number of populations & $M$ & & 1 & 1 & 100 & 300\\\hline
\hline
\end{tabular}\label{table:parameters}
\end{adjustwidth}
\end{table}

We performed numerical simulations of the microscopic, mesoscopic and macroscopic dynamics using an Euler–Maruyama scheme with time step $dt=10^{-4}s$.
In the single-population scenario considered in Figs.~\ref{fig:meta} and \ref{fig:updown}, we ran the simulations for $T_\text{sim} = 100'000s$ to obtain significant statistics.

\subsubsection*{A circular environment}
In the ring-attractor network with a single environment stored in the synaptic connectivity, we ran the simulations long enough to have at least $5'000$ burst events, which allows for a meaningful comparison of the different models.
In Table~\ref{table:1env}, we list the simulation length $T_\text{sim}$ together with an overview over the simulation results.

\begin{table}[!ht]
\begin{adjustwidth}{-0.in}{0in} 
\centering
\caption{
{\bf Simulation results complementing Fig.~\ref{fig:1Env}.}}
\begin{tabular}{|l+l|l|l|l|}
\hline
& {\bf Mesoscopic} & {\bf Microscopic} & {\bf RT model} & {\bf Macro$_{\mu=-0.9}$} \\ \thickhline
$T_\text{sim}$ [s] & 4000 & 4000 & 2350 & 2350\\ \hline
\# bursts & 5030 & 5040 & 5096 & 5167 \\
slope(\# peaks/duration) & 9.28 & 9.26 & 7.80 & 6.85 \\
slope(distance/duration) & 17.27 & 17.17 & 17.56 & 16.81 \\
mean(IBI) & 0.652 & 0.651 & 0.293 & 0.316 \\
CV(IBI) & 0.846 & 0.842 & 0.794 & 0.847  \\
skewness $\gamma_s$(IBI) & 0.917 & 0.940 & 1.230 &  1.224 \\
resc.\ skewness $\alpha_s$(IBI) & 0.361 & 0.372 & 0.516 &  0.481 \\
kurtosis $\gamma_e$(IBI) & -0.115 & -0.047 & 0.499 & 0.454 \\
resc.\ kurtosis $\alpha_e$(IBI) & -0.011 & -0.004 & 0.053 & 0.042 \\\hline
\# NLE ($>1$ peak) & 1019 & 967 & 939 & 1140 \\
fraction(NLE/bursts) & 20.3\% & 19.2\% & 18.4\% & 22.1\% \\
fraction(forward/NLE) & 51.03\% & 47.88\% & 51.54\% & 49.39\% \\
mean(abs(NLE speed)) & 12.41 & 12.54 & 12.79 & 12.64 \\\hdashline
Serial correlations &  &  &  & \\
Lag 1 (event speed) & 0.048  & 0.054  & -0.563 & -0.344 \\
Lag 1 (forward/backward) & 0.062 & 0.062 & -0.525 & -0.331 \\
Lag 2 (event speed) & -0.043 & -0.010 & 0.434 & 0.229 \\
Lag 2 (forward/backward) & -0.041 & -0.001 & 0.401 & 0.226 \\
Lag 3 (event speed) & -0.013 & 0.053 & -0.320 & -0.074 \\
Lag 3 (forward/backward) & -0.018 & 0.056 & -0.312 & -0.082 \\
Lag 4 (event speed) & -0.010 & 0.011 & 0.227 & 0.021 \\
Lag 4 (forward/backward) & -0.015 & 0.007 & 0.215 & 0.024 \\
Lag 5 (event speed) & 0.008 & -0.032 & -0.139 & 0.058 \\
Lag 5 (forward/backward) & 0.007 & -0.033 & -0.135 & 0.052 \\\hline
\hline
\end{tabular}
\begin{flushleft} We computed the statistics for the sequence of interburst intervals (IBI) $T_i, i=1,2,\dots,$ as follows: the mean is the first cumulant $\kappa_1 = \langle T_i \rangle$; the coefficient of variation (CV) is CV$=\sqrt{\kappa_2}/\kappa_1$ with second cumulant $\kappa_2 = \langle T_i^2\rangle - \kappa_1^2$; the skewness is $\gamma_s = \kappa_3 \kappa_2^{-3/2}$ with third cumulant $\kappa_3 = \langle T_i^3\rangle - 3\kappa_1\kappa_2 - \kappa_1^3$; the rescaled skewness is $\alpha_s = \gamma_s / (3CV) = \kappa_1\kappa_2/(3\kappa_2^2)$; the kurtosis is $\gamma_e = \kappa_4 \kappa_2^{-2}$ with fourth cumulant $\kappa_4 = \langle T_i^4\rangle - 4\kappa_1\kappa_3 - 3\kappa_2^2 - 6\kappa_1^2\kappa_2 - \kappa_1^4$; the rescaled kurtosis is $\alpha_e = \gamma_e / (15CV^2) = \kappa_1^2\kappa_4/(15\kappa_2^3)$. Similar to the definition of the CV, for which the Poisson process serves as a reference for the IBI variability with CV$=1$, the rescaled skewness $\alpha_s$ and rescaled kurtosis $\alpha_e$ use the inverse Gaussian distribution as a reference. Values of $\alpha_s$ larger (smaller) than $1$ and $\alpha_e$ larger (smaller) than $0$ indicate that the IBI distribution is more (less) skewed and more (less) peaked, respectively, than an inverse Gaussian, see \cite{SchFis10}.
\end{flushleft}
\label{table:1env}
\end{adjustwidth}
\end{table}

In more detail, we defined burst events as contiguous epochs of the averaged population activity above a certain threshold (taken as the activity averaged across neurons and the whole simulation period).
There is an almost perfect agreement between the microscopic network and our mesoscopic description: First, the number of bursts ($5'040$ vs.\ $5'030$) coincides up to an error of less than $0.2\%$. Second, the slopes of the linear regression between the number of peaks and the duration per burst are almost the same, see also Fig.~\ref{fig:1Env}Dii and Eii, and our model prediction ($\sim9.3$) is closer to the value observed experimentally ($\sim10.0$, \cite{DavKlo09,GupvdM10}) than the macroscopic model predictions.
Third, linear regression between the distance traveled during an event and its duration, yield almost identical slopes for the micro- and the mesoscopic models (Fig.~\ref{fig:1Env}Diii and Eiii). 
Fourth, also the means and the coefficients of variation (CV) of the interburst intervals, i.e.\ the time from the end of the $k$th burst until the start of the $k+1$st burst, coincide.
Next, we defined nonlocal replay events (NLEs) as bursts of the average activity with more than one peak so that a transient traveling wave is visible in the density plots in Fig.~\ref{fig:1Env}Bii and Cii, resembling a hippocampal replay pattern.
Among all bursts events, around $20\%$ are NLEs, which is consistent across all four models. Around half of all the NLEs travel in anti-clockwise/negative direction (``forward replay'') and the other half in clockwise/positive direction (``backward replay'' or ``preplay''); also this feature is consistent across the different models. To determine the absolute event speed per NLE, see Fig.~\ref{fig:1Env}Div and Eiv, we divided the distance of the traveled path (irrespective of in forward or backward direction) by the duration of the NLE. The means of the absolute NLE speeds in all four models were again close to each other, which stresses the robustness of our findings across parameters and models. 
Finally, we also investigated serial correlations of the NLEs. Positive serial correlations of the event speed (now taking into account also the travel direction by considering negative speed for backward replays) at lag $n$ indicate that the $k$th and the $k+n$th NLE are more likely to travel in the same direction, whereas negative correlations indicate these NLEs to travel in opposite directions.
While the macroscopic and the RT model showed strong negative correlations at lag $1$ and positive correlations at lag $2$, correlations in the mesoscopic and microscopic models were negligible, see also Fig.~\ref{fig:1EnvB}B. Besides, computing the serial correlations not for the (directional) event speed, but for a binary vector of forward/backward replay directions (by taking the sign of the directional event speed) yielded comparable results (see Table~\ref{table:1env}).

\subsubsection*{Multiple circular environments}
In the ring-attractor network with three circular environments stored in the synaptic connectivity, we ran micro- and mesoscopic simulations long enough to match the number of bursts in the deterministic (macroscopic and RT) model simulations of length $T_\text{sim} = 1'000s$, see Table~\ref{table:3env}.
To reduce confounding asymmetries in the underlying synaptic connectivity structure, we constructed the selectivity vectors $\zeta_\alpha$ in Eq.~\eqref{eq:J_3env} pseudo-randomly while guaranteeing that the number of bursts was equally distributed across the three environments, see Table~\ref{table:3env}.
More precisely, we created the binary selectivity vectors $\zeta_\alpha$ under the constraint that exactly $fM = 90$ of in total $M=300$ units $\alpha = 1,\dots,M$ are selective for each environment $k \in \{1,2,3\}$. 
Out of these $90$ selective units for environment $k$, units $1,\dots,7$ were selective for all three environments, units $8,\dots,17$ were selective for environments $k$ and $j$ and units $18,\dots,27$ for environments $k$ and $l$ with $j,l \in \{1,2,3\}, k \neq j \neq l\neq k$. The remaining $63$ units were exclusively selective for environment $k$. After this selection process, we randomly shuffled these $90$ units and drew unique, evenly distributed place field angles $\theta_\alpha^k \in \{2 \pi/90, 4\pi/90,\dots, 2\pi\} $ for each unit $\alpha = 1,\dots,90$ selective for environment $k$.

To quantify which subsequences occurred more frequent than others, we computed the probabilities for subsequences with three distinct environments $(k,j,l)\in\{1,2,3\}^3$ with $k\neq j \neq l \neq k$ by dividing the number of occurrences of a particular subsequence by the number of all possible 3-sequences (= number of all bursts$- 2$). The results are shown in Fig.~\ref{fig:3env}.

\begin{table}[!ht]
\begin{adjustwidth}{-0.in}{0in} 
\centering
\caption{
{\bf Simulation results complementing Fig.~\ref{fig:3env}.}}
\begin{tabular}{|l+l|l|l|l|}
\hline
& {\bf Mesoscopic} & {\bf Microscopic} & {\bf RT model} & {\bf Macro$_{\mu=-0.35}$} \\ \thickhline
$T_\text{sim}$ [s] & 1350 & 1350 & 1000 & 1000\\ \hline
\# bursts & 2291 & 2296 & 2354 & 2259\\
bursts in env.\ 1 & 34.57\% & 33.71\% & 32.71\% & 33.24\% \\
bursts in env.\ 2 & 31.12\% & 31.66\% & 34.45\% & 33.95\% \\
bursts in env.\ 3 & 34.31\% & 34.63\% & 32.84\% & 32.80\% \\\hline
\hline
\end{tabular}
\label{table:3env}
\end{adjustwidth}
\end{table}

\section*{Acknowledgments}
This research has received funding from the European Union’s Horizon 2020 research and innovation
programme under the Marie Sk{\l}odowska-Curie grant agreement no. 101032806 and the Swiss National Science Foundation (grant no. 200020\_184615).

\nolinenumbers

%
%
%







\newpage
\section*{Supporting Information}

\setcounter{figure}{0}

\renewcommand{\thefigure}{S\arabic{figure}}

\begin{figure}[htbp]
\centering{
\includegraphics[width=5.5in]{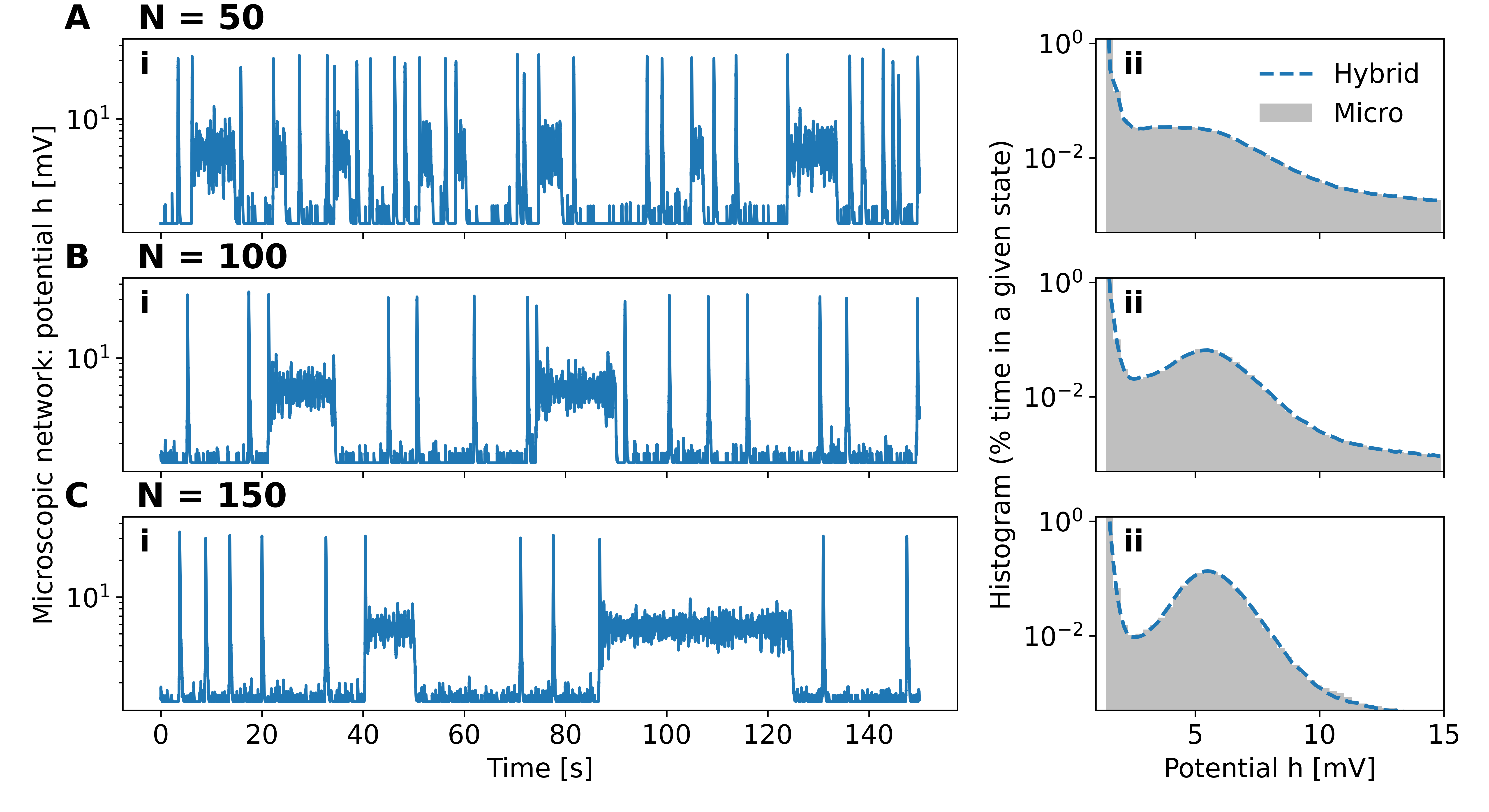}
\caption{{\bf Up-Down dynamics depend on network size.} Finite-size effects on the Up-Down dynamics in a network of (A) $N=50$, (B) $N=100$, (C) $N=150$ LNP neurons with STD. Column i: Trajectories of the input potential $h(t)$ for the microscopic network Eq.~\eqref{eq:micro} with logarithmic y-scale with the same parameters as in Fig.~\ref{fig:updown}. The larger the network size, the longer the Up states and the less frequent population spikes. Column ii: Histogram of the potential $h(t)$ shows excellent agreement between microscopic (black) and mesoscopic dynamics (blue; Eq.~\eqref{eq:model-depress-3}). Simulation length $T_\text{sim} = 10'000$~s.}
\label{fig:supp1}}
\end{figure}

\begin{figure}[htbp]
\begin{adjustwidth}{-1.0in}{-1in}
\centering
\includegraphics[width=7.5in]{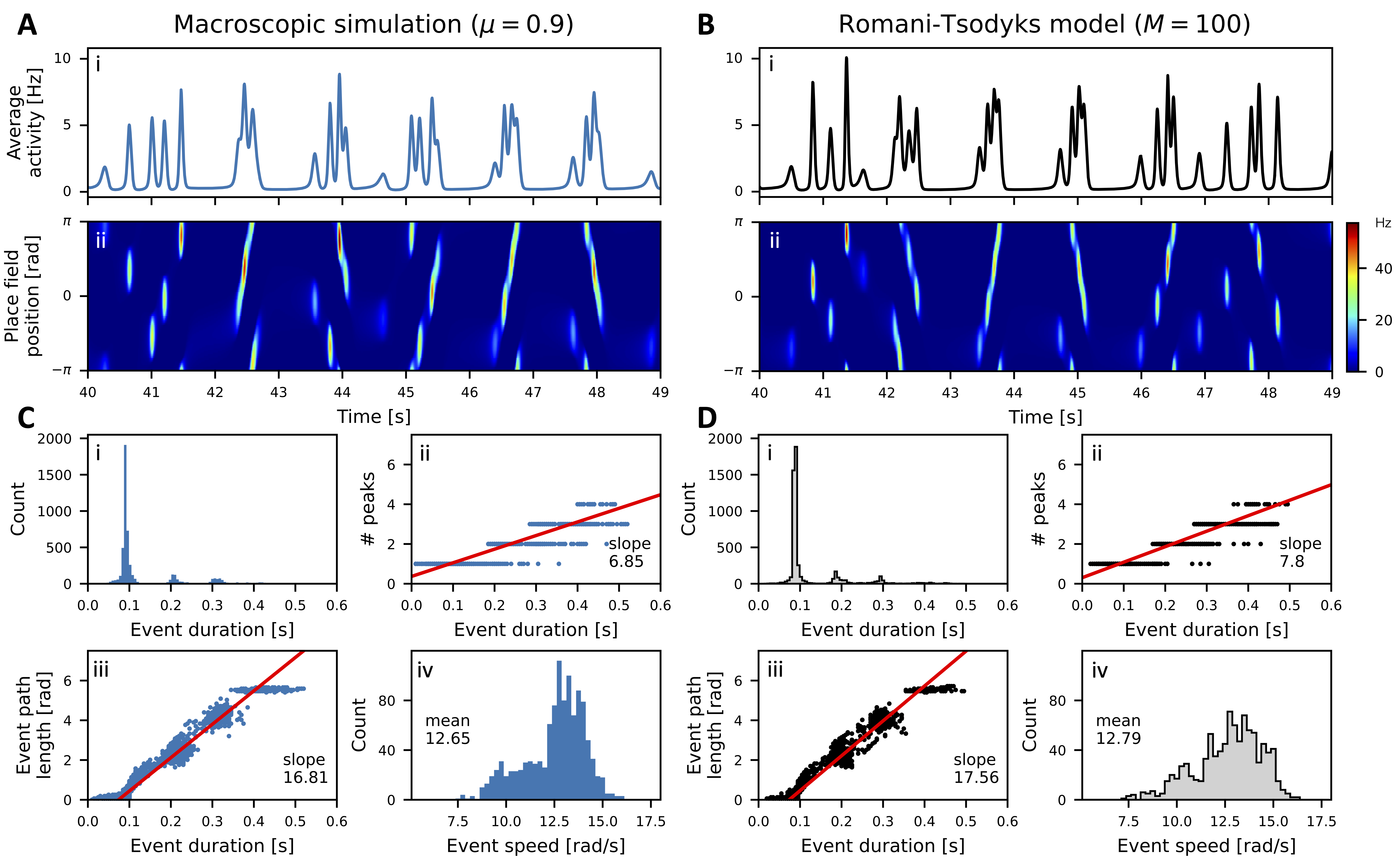}
\end{adjustwidth}
\captionsetup{width=7.5in}
\caption{{\bf Fatigue-induced hippocampal replay in the macroscopic and in the Romani-Tsodyks model.} 
Simulation results of the deterministic (macroscopic and Romani-Tsodyks) models. Panels correspond to those in Fig.~\ref{fig:1Env}.
}
\label{fig:supp2}
\end{figure}

\end{document}